\RequirePackage{snapshot}
\documentclass[fleqn,usenatbib]{mnras}

\usepackage{newtxtext,newtxmath}

\usepackage[T1]{fontenc}
\usepackage{ae,aecompl}

\usepackage{graphicx}	
\usepackage{amsmath}	
\usepackage{amssymb}	
\usepackage{morefloats}
\usepackage[dvipsnames]{xcolor}
\usepackage{bm}

\usepackage{etoolbox}
\makeatletter
\patchcmd\@combinedblfloats{\box\@outputbox}{\unvbox\@outputbox}{}{%
   \errmessage{\noexpand\@combinedblfloats could not be patched}%
}%
 \makeatother

\renewcommand{\v}[1]{\ensuremath{\mathbf{#1}}} 
\newcommand{\gv}[1]{\ensuremath{\mbox{\boldmath$ #1 $}}} 
\newcommand{\uv}[1]{\ensuremath{\mathbf{\hat{#1}}}} 
\newcommand{\pd}[2]{\frac{\partial #1}{\partial #2}} 
 
\newcommand{\grad}[1]{\gv{\nabla} #1} 
\renewcommand{\div}[1]{\gv{\nabla} \cdot #1} 
\let\baraccent=\= 
\renewcommand{\=}[1]{\stackrel{#1}{=}} 

\title[Single-scattering radiation pressure in galaxies]{Radiation pressure in galactic disks: stability, turbulence, and winds in the single-scattering limit}

\author[B. D. Wibking et al.]{
Benjamin D. Wibking,$^{1}$\thanks{E-mail: wibking.1@osu.edu}
Todd A. Thompson,$^{1}$
and Mark R. Krumholz$^{2}$
\\
$^{1}$Department of Astronomy and Center for Cosmology and AstroParticle Physics, Ohio State University,\\ 140 W 18th Ave, Columbus, OH, USA\\
$^{2}$Research School of Astronomy \& Astrophysics, Australian National University, Canberra, ACT, Australia\\
}

\date{Accepted XXX. Received YYY; in original form ZZZ}

\pubyear{2017}

\begin{document}
\label{firstpage}
\pagerange{\pageref{firstpage}--\pageref{lastpage}}
\maketitle

\begin{abstract}
The radiation force on dust grains may be dynamically important in driving turbulence and outflows in rapidly star-forming galaxies. Recent studies focus on the highly optically-thick limit relevant to the densest ultra-luminous galaxies and super star clusters, where reprocessed infrared photons provide the dominant source of electromagnetic momentum. However, even among starburst galaxies, the great majority instead lie in the so-called ``single-scattering'' limit, where the system is optically-thick to the incident starlight, but optically-thin to the re-radiated infrared. In this paper we present a stability analysis and multidimensional radiation-hydrodynamic simulations exploring the stability and dynamics of isothermal dusty gas columns in this regime. We describe our algorithm for full angle-dependent radiation transport based on the discontinuous Galerkin finite element method. For a range of near-Eddington fluxes, we show that the medium is unstable, producing convective-like motions in a turbulent atmosphere with a scale height significantly inflated compared to the gas pressure scale height and mass-weighted turbulent energy densities of $\sim 0.01-0.1$ of the midplane radiation energy density, corresponding to mass-weighted velocity dispersions of Mach number $\sim 0.5-2$. Extrapolation of our results to optical depths of $10^3$ implies maximum turbulent Mach numbers of $\sim20$. Comparing our results to galaxy-averaged observations, and subject to the approximations of our calculations, we find that radiation pressure does not contribute significantly to the effective supersonic pressure support in star-forming disks, which in general are substantially sub-Eddington. We further examine the time-averaged vertical density profiles in dynamical equilibrium and comment on implications for radiation-pressure-driven galactic winds.
\end{abstract}




\section{Introduction}

A mystery of galaxy formation is the mechanism of galactic winds, which must transport significant amounts of gas out of nearly all galaxies, as indirectly inferred from the comparison of the cosmic stellar mass function with simulations (e.g., \citealt{Somerville_2008}), the mass-metallicity relation \citep{Finlator_2008,Peeples_2011}, and chemical evolution models of the deuterium-to-hydrogen abundance ratio of the Galaxy \citep{Weinberg_2016}.

Radiation pressure on dust has been proposed as a mechanism for galactic winds by \cite{Murray_2005,MMT_2011}. Additionally, star-forming disks must be supported against collapse by the turbulent velocity dispersion of their gas, for which the driving mechanism plausibly may be gravitational instability, supernovae, or radiation pressure (e.g., \citealt{Thompson_2005,Krumholz_2016,Krumholz_2017}). On sub-galactic scales, a mechanism is sought for the observed early destruction of dense gas clumps (objects of number density $\sim 10^5$ cm$^{-3}$ and diameter $\sim 1$ pc; e.g., \citealt{Lopez_2011,Pellegrini_2011,Lopez_2014}) prior to their first supernovae.  One possible mechanism is the radiation pressure of starlight on dust \citep{Harwit_1962,Odell_1967,Chiao_1972,Barsella_1989,Ferrara_1990,Scoville_2001,Scoville_2003,Krumholz_2009,MQT_2010,Raskutti_2016}.  In this work, we study aspects of the radiation pressure mechanism, with application to these dynamical questions.

Previous work on radiation pressure-driven winds and turbulence in the galactic context has largely focused on the radiative force imparted by IR photons in IR-optically thick dusty gas columns (\citealt{Krumholz_2012,Krumholz_2013,Davis_2014,Zhang_2017}; but see \citealt{Raskutti_2016,Tsang_2017}), where multiple scattering of IR photons can transfer many times the photon source momentum $L/c$ to the gas. This ``multiple-scattering'' regime applies to systems (e.g., galaxies or GMCs) with very high column densities ($0.1$--$10 \, \text{g} \, \text{cm}^{-2}$ or $10^3$--$10^4 \, M_{\sun} \, \text{pc}^{-2}$) (\citealt{Thompson_2005}; \citealt{Andrews_2011}), but is inapplicable to systems that have gas column densities and dust-to-gas ratios similar to that of the Galactic interstellar medium or an `average' local starburst galaxy like M82 \citep{Coker_2013}.

This latter regime is that of ``single-scattering'' of UV/optical photons from starlight; that is, systems in which the optical depth to scattering of UV/optical photons by dust is of order unity, but the optical depth to scattering of IR photons by dust is much less than unity.  This regime is applicable to a wider variety of systems because the flux-mean opacity $\kappa_F = \int \kappa_{\nu} F_{\nu} d\nu / \int F_{\nu} d\nu$ integrated over a galaxy's starlight SED is $2-3$ orders of magnitude larger than the opacity integrated over the IR band alone \citep{Andrews_2011,Thompson_2005,Draine_2011} (although expanding IR-optically thick media must eventually undergo a single-scattering phase; \citealt{Thompson_2015}). High redshift star-forming galaxies also have column densities in the single-scattering range \citep{Bouche_2007,Daddi_2008,Daddi_2010,Genzel_2010,Tacconi_2013}; thus, this limit is broadly applicable.

This paper focuses on the ``single-scattering'' regime of direct radiation from starlight onto dusty gas. We investigate the nonlinear instabilities and dynamics in this regime with multidimensional simulations of the astrophysically-relevant limit of compressible radiation hydrodynamics.\footnote{For an investigation of linear instability in the incompressible hydrodynamic regime with radiation forces, see \cite{Krolik_1977}.} We conduct simulations in an idealized 2D plane-parallel geometry of a perfectly coupled isothermal dust-gas mixture subject to both gravitational forces and radiation forces and measure the resulting turbulent velocity dispersion and kinetic energy density. We investigate the general stability properties of the medium, whether radiation pressure can drive turbulence, and if so, how much, and illuminate regimes where single-scattering radiation pressure might be dynamically important.

In section \ref{section:perturbation_theory}, we derive a hydrostatic equilibrium profile in the presence of gravity and radiation pressure in the single-scattering limit and investigate its linear stability. In section \ref{section:methods}, we describe our numerical methods for hydrodynamics and radiation transport, and our initial conditions and boundary conditions. In section \ref{section:stability}, we describe the results of our numerical experiments. We assess the stability of sub-Eddington atmospheres in the single-scattering limit over a broad range of parameter space and compute the resulting turbulent velocity dispersion, turbulent energy density, and vertical density profiles in the unstable cases. We discuss the implications for driving mechanisms of galactic turbulence and winds in section \ref{section:discussion} and conclude in section \ref{section:conclusion}.

\section{Linear perturbation analysis}
\label{section:perturbation_theory}
We consider a radiation-supported atmosphere with an analytic model of radiation-hydrostatic equilibrium, in which pressure gradients, gravitational forces, and radiation forces produce a time-stationary state with zero velocity.\footnote{Previous works have derived qualitatively similar results (e.g., density profile inversions) in the different physical regime of local thermodynamic equilibrium (e.g., \citealt{Joss_1973}), which does not apply to the single-scattering case.}

For an isothermal\footnote{Isothermal is an appropriate assumption about the thermal state of the gas because for most systems in the single-scattering regime, the gas density is low enough ($\lesssim 10^{4.5}$ cm$^{-3}$) that the dust and gas are not thermally coupled.} gravitating atmosphere with sound speed $c_T$ and gravitational acceleration $-g$, we have the exponential density profile:
\begin{align}
\rho(z) = \rho_0 e^{-z/h_0}
\end{align}
where $h_0 = c_T^2 /g$ is the scale height of the atmosphere, $c_T$ is the isothermal sound speed, and $\rho_0$ is a density normalization that sets the total column density of the atmosphere.

In the optically-thin limit, this is modified by replacing $-g$ with the net acceleration on a parcel of gas due to both radiation forces and gravity.  For zero net acceleration, there is no unique equilibrium profile and for positive net acceleration, there is no equilibrium state at all.
  
In the case of perfectly beamed radiation from an infinite midplane source, the equations for the equilibrium density profile are
\begin{align}
\frac{d\rho}{dz} &= \frac{1}{c_T^2} \left(-g + \frac{\kappa}{c} F_{\text{mid}} e^{-\tau} \right) \rho(z) \, , \text{ and} \\
\frac{d\tau}{dz} &= \kappa \rho(z) \, ,
\end{align}
where $\rho_0$ is the gas density at the midplane, $\tau$ is the vertical optical depth, $F_{\text{mid}}$ is the (beamed) radiation flux at the midplane, and $\kappa$ is the flux-mean dust opacity per mass of dust-gas mixture. The boundary conditions are
\begin{align}
\rho(0) &= \rho_0 \, , \\
\tau(0) &= 0 \, , \text{ and} \\
F_{\text{mid}} &= \Gamma_{\text{Edd}} F_{\text{Edd,beamed}}(\tau) \, .
\end{align}

We compute the beamed Eddington flux by solving for the mass-weighted radiative acceleration
\begin{equation}
F_{\text{Edd,beamed}}(\tau) = \frac{g c}{\kappa} \frac{\tau}{1 - e^{-\tau}}
\end{equation}
that produces an acceleration equal to $g$. We compute a hydrostatic profile for a given Eddington ratio $\Gamma_{\text{Edd}}$ and optical depth $\tau$ by solving for the value of $\rho_0$ that yields a profile of desired optical depth $\tau$. The density $\rho$ is positive whenever $\Gamma_{\text{Edd}} < 1$, and unphysical solutions are obtained for $\Gamma_{\text{Edd}} \ge 1$.

\begin{figure}
  \includegraphics[width=\columnwidth]{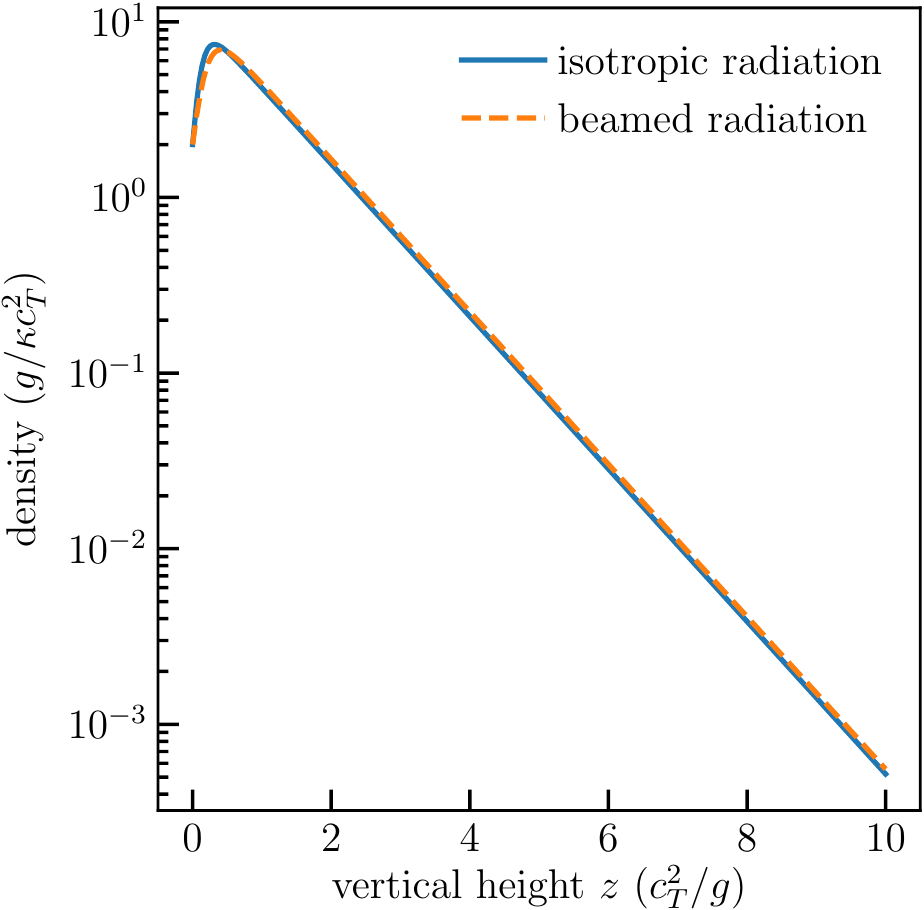}
  \caption{Hydrostatic density profiles with $\Gamma_{\text{Edd}}=0.8$ and $\tau=10$ computed on the interval $0 < z < 10$ with 512 grid points. The dashed orange line shows the hydrostatic profile assuming a source of beamed radiation. The solid blue line shows the hydrostatic profile assuming an isotropic radiation source.}
  \label{fig:hse}
\end{figure}

We show a numerically-integrated hydrostatic profile for $\Gamma_{\text{Edd}} = 0.8$ and optical depth $\tau = 10$ in Figure \ref{fig:hse}. We note that a density profile inversion develops when $F_{\text{mid}} > g c / \kappa$, as is the case for the `beamed radiation' profile shown in Figure \ref{fig:hse}.

In the case of an isotropically-emitting infinite midplane source, the equations for the equilibrium density profile are
\begin{align}
\frac{d\rho}{dz} &= \frac{1}{c_T^2} \left( -g + \frac{\kappa}{c} F_{\text{mid}} 2 \int_{0}^{1} e^{-\tau(z) / \mu} \, \mu \, d\mu \right) \, \rho(z) \, , \text{ and}\\
\frac{d\tau}{dz} &= \kappa \rho(z) \, ,
\end{align}
where $F_{\text{mid}}$ is the (isotropic) flux at the midplane, $\tau$ is the vertical optical depth, and $\mu$ is the direction cosine with respect to the vertical $z$ axis.

The boundary conditions are:
\begin{align}
\rho(0) &= \rho_0 \, , \\
\tau(0) &= 0 \, , \text{ and} \\
F_{\text{mid}} &= \Gamma_{\text{Edd}} F_{\text{Edd,isotropic}}(\tau) \, .
\end{align}
To derive $F_{\text{Edd,isotropic}}(\tau)$, we compute the flux which produces a mass-weighted radiative acceleration equal to $g$. Using Chandrasekhar's definition of the flux \citep{Chandrasekhar_1960} and assuming gray radiation, we have
\begin{equation}
\pi F_{\hat{z}} = \int_{S^2} I (\hat{n}) \, \hat{n} \cdot \hat{z} \, d \Omega \, ,
\label{eq:chandrasekhar_flux}
\end{equation}
where $d\Omega$ is the element of solid angle $\sin \phi\, d\theta\, d\phi$, and $\mu = \cos^{-1} (\hat{n} \cdot \hat{z})$ is the angle with respect to the $\hat z$ direction for which we wish to compute the flux. In spherical polar coordinates, we have
\begin{equation}
\pi F(\tau) = 2 \pi I_{0} \int_{0}^{\pi / 2} e^{-\tau / \cos{\phi}} \cos{\phi} \sin{\phi} \, d\phi
\end{equation}
\begin{equation}
\pi F(\tau) = 2 \pi I_{0} \int_{0}^{1} e^{-\tau / \mu} \mu \, d\mu
\end{equation}
\begin{equation}
\pi F(\tau) = \pi I_{0} e^{-\tau} \left( 1 - \tau - \tau^2 e^{\tau} \int_{\tau}^{\infty} e^{-t}/t \, dt \right) \, ,
\end{equation}
where $I_0$ is the (isotropic) specific intensity at $z=0$.

To obtain the mass-weighted Eddington flux, we need to compute the mass-weighted mean radiative acceleration and solve for the flux at $z=0$ that produces such an acceleration. Since the opacity $\kappa$ is constant and the radiative acceleration has a prefactor $\kappa \rho$ that multiplies the flux, we can perform this weighting by integrating the flux over optical depth:
\begin{equation}
\begin{split}
\int_{0}^{\tau} F(\tau') \, d\tau' = I_{0} \int_{0}^{\tau} e^{-\tau'} \left( 1 - \tau' - \tau'^2 e^{\tau'} \int_{\tau'}^{\infty} e^{-t}/t \, dt \right) \, d\tau'\\
= \frac{2}{3} I_{0} \left[ 1 + e^{-\tau} \left(\frac{\tau}{2} - \frac{\tau^{2}}{2} - 1\right) + \frac{\tau^3}{2} \int_{\tau}^{\infty} \frac{e^{-t}}{t} \, dt \right]
\end{split}
\end{equation}
to find that the plane-parallel isotropic Eddington flux is
\begin{equation}
\begin{split}
F_{\text{Edd,isotropic}}&(\tau) = \frac{3}{2} \frac{g c}{\kappa} \tau \\
&\times \, \left[ 1 + e^{-\tau} \left(\frac{\tau}{2} - \frac{\tau^{2}}{2} - 1\right) + \frac{\tau^3}{2} \int_{\tau}^{\infty} \frac{e^{-t}}{t} \, dt \right]^{-1} .
\label{eq:fedd_ss}
\end{split}
\end{equation}

As before, $\rho_0$ is determined by solving for the value that yields a profile with the desired optical depth $\tau$.  The density $\rho$ is found to be positive for $\Gamma_{\text{Edd}} < 1$, and unphysical solutions are obtained for $\Gamma_{\text{Edd}} \ge 1$.  We show a numerical integration for a hydrostatic density profile where $\Gamma_{\text{Edd}} = 0.8$ and $\tau = 10$ in Figure \ref{fig:hse}.  A density inversion develops whenever $F_{\text{mid}} > g c / \kappa$, as is the case for the `isotropic radiation' profile shown in Figure \ref{fig:hse}.

\subsection{Perturbations}
We now perturb the hydrostatic density profile in order to investigate its linear stability. In this section we assume that the radiation from the midplane is perfectly beamed, which is not physically realistic and is not used in our main set of simulations.  We make this assumption here only because it enables us to treat the radiative transfer consistently at linear order within the perturbation analysis that follows.

We linearize the equations as
\begin{align}
\pd{\delta \rho}{t} + \rho_0 \div{\delta \v{v}} = 0 \, , \text{ and} \\
\rho_0 \pd{\v{\delta v}}{t} = - \grad P + \rho \v{g} + \rho \frac{\kappa \v{F}_{\text{rad}}}{c} \, ,
\end{align}
where $\rho = \rho_0 + \delta \rho$, $P = P_0 + \delta P$, and $F_{\text{rad}} = F_{\text{rad},0} + \delta F_{\text{rad}}$ refer to the sum of the hydrostatic background state and the perturbed state. Subtracting the hydrostatic background state and dropping second-order terms, we have
\begin{equation}
\rho_0 \pd{\v{\delta v}}{t} = - c_T^2 \grad \delta\rho + \delta\rho \v{g} + \rho_0 \frac{\kappa \v{\delta F}_{\text{rad}}}{c} + \delta\rho \frac{\kappa \v{F}_{\text{rad},0}}{c} \, .
\end{equation}
With background state optical depth $\tau_0$ and the perturbed optical depth $\delta \tau$,
\begin{align}
F(z) &= F_{\text{mid}} \, e^{-(\tau_0 + \delta \tau)} = F_{\text{mid}} \, e^{-\tau_0} \, e^{-\delta \tau} \, , \\
F_0(z) &= F_{\text{mid}} \, e^{-\tau_0} \, , \text{ and} \\
\delta F(z) &= F_0 ( e^{-\delta \tau} - 1 ) \approx F_0 ( -\delta \tau ) \, .
\end{align}
We then assume solutions of the form 
\begin{equation}
\delta \v{v}(x,z,t) = \delta \v{v} \, e^{i k_x x + i k_z z - i \omega t} \, ,
\label{eq:wkb}
\end{equation}
compute
\begin{equation}
\delta \tau = \int_0^z \kappa \delta\rho(z) dz' = \frac{\kappa \delta\rho}{-i k_z} \, ,
\end{equation}
and use the perturbed continuity equation to obtain
\begin{equation}
\frac{-i \omega^2}{\v{k} \cdot \v{\delta v}} \delta \v{v} = -i \v{k} c_T^2 + \v{g} + \frac{\kappa}{c} \v{F_0} \left( 1 - \frac{\kappa \rho_0}{i k_z} \right) \, .
\end{equation}
With further algebra we obtain a local dispersion relation
\begin{equation}
\omega^2 = \left(k_x^2 + k_z^2\right) c_T^2 + i k_z g + \frac{\kappa}{c} F_0(z) \left[i k_z - \kappa \rho_0(z)\right] \, ,
\end{equation}
which is always stable for $k_z > 0$, and is unstable for $k_z = 0$ whenever the horizontal wavenumber $k_x$ is less than a critical wavenumber
\begin{equation}
k_{x,\text{c}} = c_T^{-1} \sqrt{ \frac{\kappa}{c} F_0(z) \, \kappa \rho_0(z) } \, .
\end{equation}
Therefore radiation-supported atmospheres with beamed radiation are unstable to perturbations with a horizontal wavelength longer than
\begin{align}
\lambda_{\text{c}} &= 2\pi / k_{x,\text{c}} = 2\pi c_T \left( \frac{\kappa}{c} F_0(z) \, \kappa \rho_0(z) \right)^{-1/2} \\
&= 4.8 \text{ pc} \, \left( \Gamma_{\text{Edd}}^{-1/2} \, \tau^{-1} \right) \, \left( \frac{T}{300 \text{ K}} \right) \left( \frac{\Sigma}{100 \, M_{\odot} \text{ pc}^{-2}} \right)^{-1} \, ,
\end{align}
in the limit $\tau \rightarrow \infty$ and assuming $\kappa \rho_0 = d\tau/dz \sim \tau \, (c_T^2/g)^{-1}$.

However, since the most unstable modes are at arbitrarily large wavelengths, the validity of the WKB approximation may be in doubt. Therefore, we carry out a quasi-global numerical eigenmode analysis with a finite difference discretization, with $\delta \v{v}$ allowed to be an arbitrary (differentiable) function of $z$ and with an $x$-dependence restricted to functions of the WKB form (i.e., $\delta \v{v} = \delta \v{v}(z) \, e^{i k_x x - i \omega t}$). Under these assumptions, we obtain a one-dimensional wave equation for $\delta v_z$ as a function of $k_x$. We then use the fact that $\partial^2 / \partial t^2 = -\omega^2$ to obtain an eigenvalue equation of the form
\begin{equation}
\bm{A} \, \v{\delta v_z} = \lambda \, \v{\delta v_z} \, ,
\end{equation}
where $\bm{A}$ is the finite difference matrix, $\lambda$ is the eigenvalue (equal to $\omega^2 - k_x^2 c_T^2$), and $\v{\delta v_z}$ is the eigenvector. We solve this eigenvalue equation as a banded diagonal matrix eigenvalue problem with \textsc{LAPACK} (\citealt{lapack}, via the \textsc{Python} interface provided by \textsc{SciPy}; \citealt{scipy}), yielding the entire set of eigenvalues and eigenvectors that can be represented on the discrete $z$-grid. We assess the fidelity of the solution of the discrete problem to that of the continuous problem by both successively doubling the resolution of the grid in $z$ and doubling the box height, finding that the lowest-order eigenmodes are affected by much less than one percent for changes about the values used here.

\begin{figure}
  \includegraphics[width=\columnwidth]{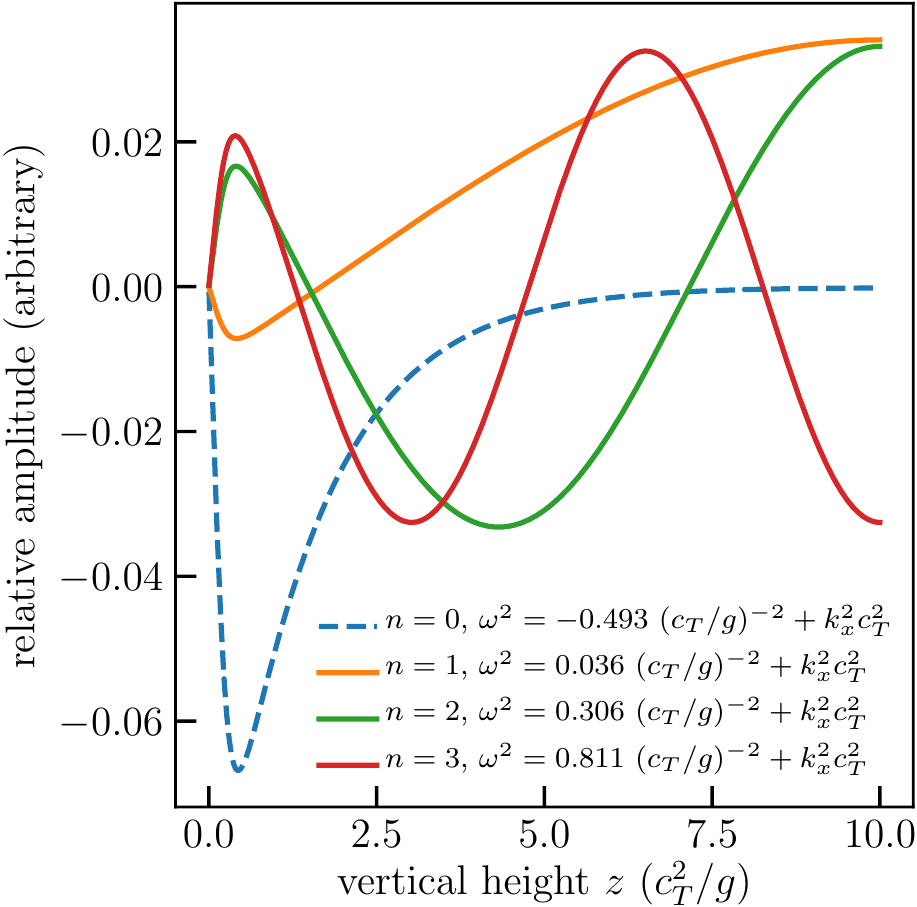}
  \caption{The lowest-order vertical eigenmodes for a hydrostatic profile with $\Gamma_{\text{Edd}}=0.8$ and $\tau=10$ computed on the interval $0 < z < 10$ with 2048 grid points. From the eigenvalue of the fundamental mode, we find that the minimum unstable horizontal wavelength $\lambda_{x,c} = 8.95 \, c_T^2/g$. The $n > 0$ modes do not converge with increasing box height, suggesting there exists a continuum of modes above the fundamental mode.}
  \label{fig:eigenmodes}
\end{figure}

Carrying out this analysis, we find the fundamental vertical eigenmode can be either stable or unstable, depending on the $(\Gamma_{\text{Edd}}, \tau)$ parameters. We likewise find that the fundamental mode is unstable only for $k_x < k_{x,c}$ for some $k_{x,c}$ that is a function of $\Gamma_{\text{Edd}}$ and $\tau$. All higher-order modes appear to be stable across the entire parameter space. We show the results of an eigenmode calculation at $\Gamma_{\text{Edd}}=0.8$, $\tau=10$ in Figure \ref{fig:eigenmodes}, where the fundamental mode is unstable for horizontal wavelengths $\lambda_{x,c} > 8.95 \, c_T^2 / g$.

\begin{figure}
  \includegraphics[width=\columnwidth]{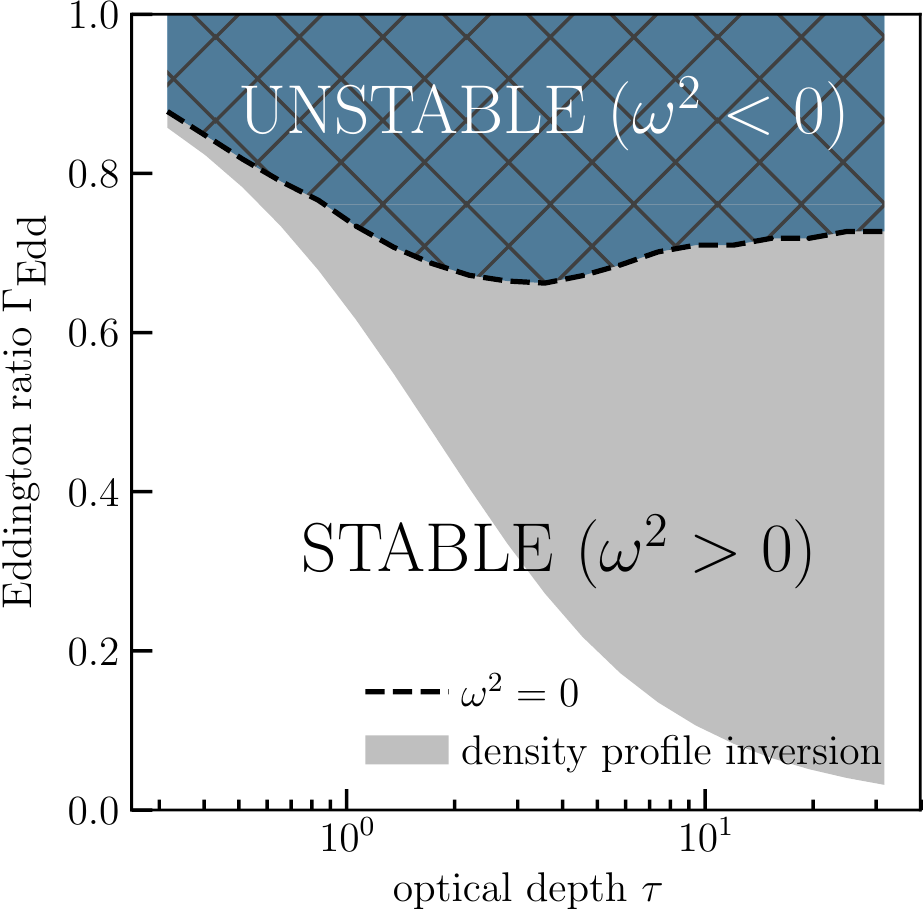}
  \caption{The solution to the equation $\omega^2(\Gamma_{\text{Edd}}, \tau) = 0$, showing the boundary between stability and instability as a function of $\Gamma_{\text{Edd}}$ and $\tau$ for beamed radiation that is predicted by our linear perturbation analysis (see section \ref{section:perturbation_theory}). At each optical depth, we solve for the critical value of $\Gamma_{\text{Edd}}$ with a bisection root finding method that iterates over the output of our eigenmode solver.  The eigenmodes are solved by finite differences on the domain $0 < z < 40$ on a grid of 1024 points.}
  \label{fig:perturbative_stability}
\end{figure}
We use our numerical eigenmode solver to solve the equation $\omega^2(\Gamma_{\text{Edd}},\tau) = 0$ and show this in Figure \ref{fig:perturbative_stability}. This represents the prediction of the boundary between stability and instability as a function of optical depth and Eddington ratio, since $\omega^2 > 0$ indicates stability and $\omega^2 < 0$ indicates instability (according to our sign convention given in eq. \ref{eq:wkb}).

In section \ref{section:stability}, we examine the agreement of this perturbative prediction with our fully nonlinear simulations. In that set of simulations, we choose an amplitude of turbulent driving that generates transonic velocity dispersions, since any representative region of a galaxy's interstellar medium will involve density perturbations of at least order unity.  However, we conduct limited tests with small-amplitude perturbations (of order $10^{-2} \, c_T$ in velocity amplitude) generated from a $k^{-4}$ power spectrum, which ensures that the perturbations are primarily at long wavelengths, as we have found those to be the most unstable. In these tests, we do not observe any instability. Since our linear perturbation analysis suggests that there should be unstable behavior in some parameter regimes, we speculate that this lack of instability in our numerical simulations with small-amplitude perturbations is due to numerical dissipation or inconsistency in our boundary conditions. This behavior may also be due to our low-order coupling between radiation and hydrodynamics (i.e., operator splitting; see section \ref{section:methods}).

\section{Numerical methods}
\label{section:methods}
Having found that radiation pressure-supported atmospheres in the single-scattering limit are unstable above a critical Eddington ratio (Figure \ref{fig:perturbative_stability}), we now turn to numerical simulations to investigate the non-linear development of the instability. We use \textsc{Athena}, a compressible Godunov code (\citealt{Stone_2008}; \citealt{Stone_2010}), in order to evolve the equations of two-dimensional isothermal hydrodynamics.  We modify the code to include a radiation force term in the momentum equation by first-order operator splitting. For the hydrodynamics, we use the second-order van Leer integrator \citep{Stone_2009} with piecewise-parabolic (PPM) interface reconstruction in the primitive variables and the HLLC Riemann solver to compute the fluxes. Due to the difficult flow conditions that we encounter, we adaptively reduce the order of the reconstruction to first-order whenever unphysical states would result from PPM reconstruction (i.e., negative densities), when the momentum source term is strongly impulsive (i.e., at interfaces where the gradient of the radiative acceleration $> 10$), or would otherwise produce unphysically large velocities ($\mathcal{M} > 100$). These conditions are regularly produced in the presence of an operator-split radiative acceleration source term at sharp optical depth gradients, and we cannot evolve our simulations in a stable manner at a reasonable timestep without resorting to such measures.\footnote{We use a CFL number (defined here as the ratio of the timestep to the crossing time of a sound wave across one grid zone along any coordinate axis in the instantaneous comoving fluid frame; see \citealt{CFL} for the original definition) of 0.4 for all of our simulations in this paper.} We use these modified criteria with the `first-order flux correction' option implemented in \textsc{Athena}.

We compute the radiation source terms by solution of the time-independent gray radiation transport equation, with angles discretized via the method of discrete ordinates, quadratures chosen to be appropriate to the two-dimensional, plane-parallel geometry (Appendix \ref{appendix:angles}), and spatial terms in the transport equation discretized via the discontinuous Galerkin method (Appendix \ref{appendix:DG}).  We discuss the dependence of our results on spatial and angular resolution and box size in Appendix \ref{appendix:resolution}.

The equations we solve are
\begin{align}
\pd{\rho}{t} + \div{\rho \v{v}} = 0 \, , \label{eq:continuity}\\
\pd{(\rho \v{v})}{t} + \div{( P_{\text{gas}} + \rho \v{v}^2 )} = (\kappa \rho / c) \v{F_{\text{rad}}} - \rho g \uv{y} \, , \\
\hat{n} \cdot \vec{\nabla} I_{\text{rad}} = - (\kappa \rho) I_{\text{rad}} \, , \\
\pi \v{F_{\text{rad}}} = \int I_{\text{rad}}(\hat{n}) \, \hat{n} \, d\Omega \, ,
\label{eq:flux}
\end{align}
and an isothermal equation of state
\begin{equation}
P_{\text{gas}} = \rho c_T^2,
\end{equation}
where we have a direction cosine $\hat{n}$, constant isothermal sound speed $c_T$, constant gravitational acceleration $g$, and a constant, temperature-independent opacity $\kappa$. The flux is defined to include the factor of $\pi$ on the left-hand side so that the relationship between intensity and flux is exactly the same for an isotropic angular distribution of radiation (eqs. \ref{eq:chandrasekhar_flux}--\ref{eq:fedd_ss}) as for a beamed angular distribution of radiation. With constant gravity, opacity, and equation of state, the equations are self-similar and the scales of the problem are set by the sound speed, the magnitude of the gravitational acceleration, and the opacity.  We therefore introduce dimensionless equations by setting the physical constants (including the speed of light $c$) to unity and switching to dimensionless density $D$, dimensionless velocity $V$, dimensionless time $T$, dimensionless intensity $I$, dimensionless flux $F$, and dimensionless energy density $E$. In these variables, equations \ref{eq:continuity}--\ref{eq:flux} become
\begin{align}
\pd{D}{T} + \div{D} = 0 \, , \\
\pd{(D \v{V})}{T} + \div{[D ( 1 + \v{V}^2 )]} =  D \v{F} - D \uv{y} \, , \\
\hat{n} \cdot \vec{\nabla} I = - D I \, , \text{ and} \\
\pi \v{F} = \int I(\hat{n}) \, \hat{n} \, d\Omega \, .
\end{align}
For the hydrodynamics equations, we impose boundary conditions that are reflecting on the lower horizontal boundary and either reflecting or outflow (with inflow disallowed by a switch, i.e., diode boundary conditions) on the upper boundary.  The horizontal boundary conditions are periodic.  For the radiation transport equation, we impose an isotropically-radiating lower boundary of fixed flux $F_{\text{mid}}$, the physical counterpart of which is the interstellar radiation field from a continuously star-forming stellar population.  We likewise impose periodic boundary conditions in the horizontal direction for the radiation transport.

The characteristic length, time, velocity, density, intensity, flux, and energy density scales are
\begin{align}
x_0 &= y_0 = \frac{c_T^2}{g} = 0.92 \, \text{pc} \left( \frac{T}{300 \,K} \right) \left( \frac{\Sigma}{100 M_{\sun} \text{pc}^{-2}} \right)^{-1} \, , \\ \label{eq:dimensionless}
t_0 &= \frac{c_T}{g} = 0.57 \, \text{Myr} \left( \frac{T}{300 \,K} \right)^{1/2} \left( \frac{\Sigma}{100 M_{\sun} \text{pc}^{-2}} \right)^{-1} \, , \\
v_0 &= c_T = 1.6 \, \text{km} \, \text{s}^{-1} \left( \frac{T}{300 \,K} \right)^{1/2} \, , \\
\rho_0 &= \frac{g}{\kappa c_T^2} \nonumber \\
&= 210 \, \text{cm}^{-3} \, \text{H} \, \left( \frac{\kappa}{10^3 \, \text{cm}^2 \, \text{g}^{-1}} \right)^{-1} \left( \frac{T}{300 \,K} \right)^{-1} \left( \frac{\Sigma}{100 M_{\sun} \text{pc}^{-2}} \right) \, , \\
I_0 &= \frac{g c}{\kappa} \nonumber \\
&= 0.263 \, \text{ergs} \, \text{cm}^{-2} \, \text{s}^{-1} \left( \frac{\kappa}{10^3 \, \text{cm}^2 \, \text{g}^{-1}} \right)^{-1} \left( \frac{\Sigma}{100 M_{\sun} \text{pc}^{-2}} \right) \, , \\
F_0 &= \frac{g c}{\kappa} \nonumber \\
&= 0.263 \, \text{ergs} \, \text{cm}^{-2} \, \text{s}^{-1} \left( \frac{\kappa}{10^3 \, \text{cm}^2 \, \text{g}^{-1}} \right)^{-1} \left( \frac{\Sigma}{100 M_{\sun} \text{pc}^{-2}} \right) \, , \label{eq:dimensionless_flux}
\end{align}
and
\begin{equation}
\begin{split}
e_{0} = \frac{2 \pi}{c} \, \frac{g c}{\kappa} = 1040 \, G_0 \left( \frac{\kappa}{10^3 \, \text{cm}^2 \, \text{g}^{-1}} \right)^{-1} \left( \frac{\Sigma}{100 M_{\sun} \text{pc}^{-2}} \right) \, , \label{eq:dimensionless_erad}
\end{split}
\end{equation}
respectively, where we have normalized $\kappa$ to an appropriate value for the flux-mean dust opacity per gram of gas for a zero-age main sequence fully populated stellar IMF and a Galactic dust-to-gas ratio. The quantity $G_0 = 5.29 \times 10^{-14}$ ergs cm$^{-3}$ is the fiducial solar neighborhood value of the interstellar radiation field (ISRF) in the $6-13.6$ eV band (\citealt{Draine_2011}; \citealt{Habing_1968}). Note that this value varies dramatically with galactocentric radius; e.g., in the central molecular zone of the Galaxy, \cite{Lis_2001} inferred an ISRF of $\sim 10^3 \, G_0$. We have assumed that the gravitational acceleration $g$ and the total surface density of mass $\Sigma$ are related by
\begin{equation}
g = 2 \pi G \Sigma = 8.7 \times 10^{-9} \, \text{cm s}^{-2} \left( \frac{\Sigma}{100 M_{\sun} \text{pc}^{-2}} \right),
\end{equation}
as appropriate for a geometrically thin disk. Note that due to the factor of $\pi$ included in the definition of the flux we adopt (eq. \ref{eq:flux}, which is identical to eq. 7 in \citealt{Chandrasekhar_1960}), our units of intensity and flux are identical.

With these dimensionless variables, the only free parameters of our simulation are the vertical optical depth $\tau$ and the single-scattering Eddington ratio $\Gamma_{\text{Edd}}$.  We must also choose the horizontal and vertical simulation box sizes $X_{\text{max}}$ and $Y_{\text{max}}$, respectively.

The vertical optical depth $\tau(X)$ of the sightline at horizontal position $X$ is
\begin{equation}
\tau(X) = \int_{0}^{Y_{\text{max}}} D(X,Y) \, dY.
\end{equation}
We denote the mean vertical optical depth averaged over all vertical sightlines (or that of a uniform vertical density profile) as $\tau$ (without any arguments):
\begin{equation}
\tau = \frac{1}{X_{\text{max}}} \int_{0}^{X_{\text{max}}} \int_{0}^{Y_{\text{max}}} D(X,Y) \, dY dX.
\end{equation}
The definition implies that the vertical optical depth of an exponential vertical density profile of the form $\rho(y) = \rho_0 \exp(-y/y_0)$ is
\begin{equation}
\tau = \rho_0 y_0
\end{equation}
for an infinitely-tall box. Note that regardless of the form of the density profile,
\begin{equation}
\tau(X) = \kappa \Sigma_{\text{gas}}(X) \, ,
\end{equation}
and averaged over all columns
\begin{equation}
\tau = \kappa \langle \Sigma_{\text{gas}} \rangle.
\end{equation}

\begin{figure}
  \includegraphics[width=\columnwidth]{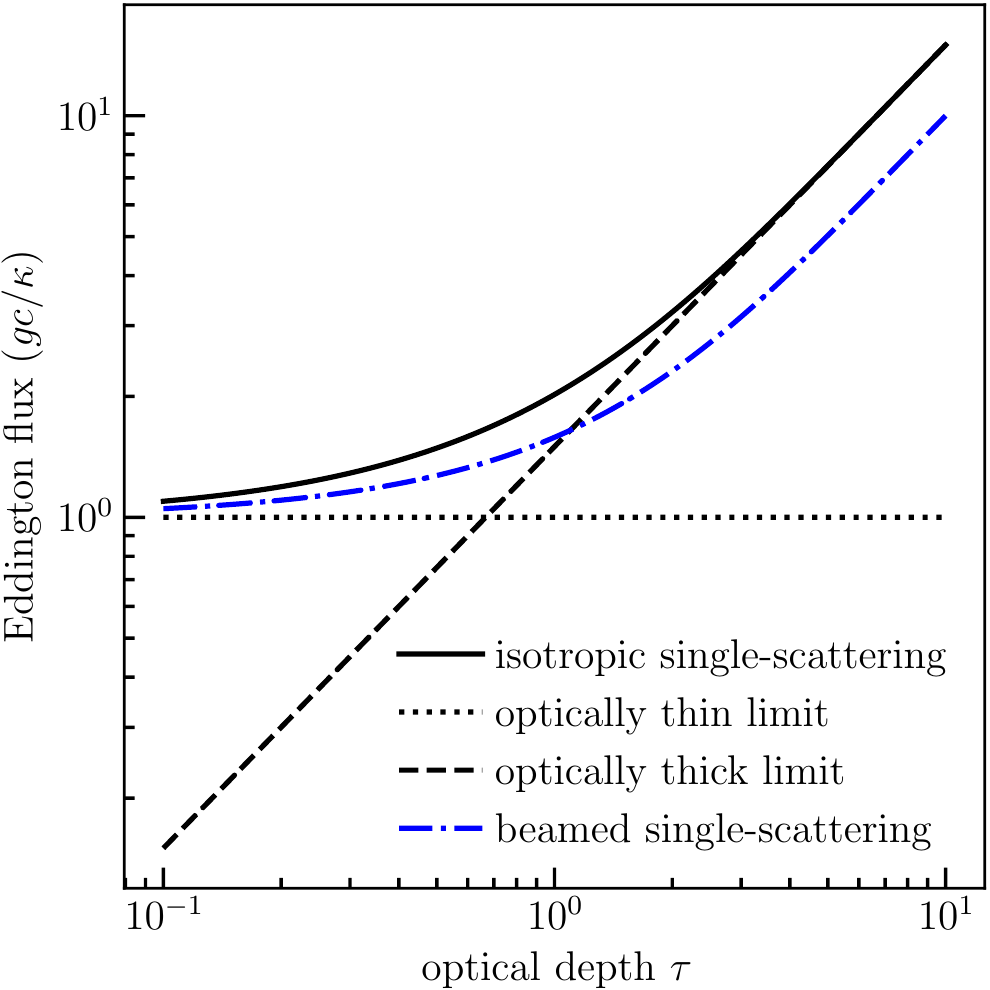}
  \caption{The various definitions of the Eddington flux (normalized to the optically-thin Eddington flux $gc/\kappa$) as a function of optical depth. The optically-thin Eddington flux is the conventional Eddington flux $gc/\kappa$ (eq. \ref{eq:fedd_thin}).  The isotropic single-scattering Eddington flux is defined in eq. \ref{eq:fedd_ss}. The optically-thick isotropic Eddington flux is given in eq. \ref{eq:fedd_thick}. The beamed single-scattering Eddington flux is defined in eq. \ref{eq:fedd_beamed}.}
  \label{fig:eddington_flux}
\end{figure}

The single-scattering Eddington ratio
\begin{equation}
\Gamma_{\text{Edd}} = \frac{F_{\text{mid}}}{F_{\text{Edd}}}
\label{eq:eddington_ratio}
\end{equation}
is defined as the ratio of the incident lower boundary flux $F_{\text{mid}}$ to the single-scattering Eddington flux $F_{\text{Edd}}$, where the single-scattering Eddington flux is given by eq. \ref{eq:fedd_ss} as derived in section \ref{section:perturbation_theory}. This is precisely the flux that produces a mass-weighted mean radiative acceleration equal and opposite that of the gravitational acceleration $g$. In the infinitely optically thick limit ($\tau \rightarrow \infty$), this becomes
\begin{equation}
F_{\text{Edd}}(\tau \rightarrow \infty) = \frac{3}{2} \frac{g c}{\kappa} \tau = \frac{3}{2} gc \langle \Sigma_{\text{gas}} \rangle \, ,
\label{eq:fedd_thick}
\end{equation}
while in the optically-thin limit we recover the classical Eddington ratio
\begin{equation}
F_{\text{Edd}}(\tau \rightarrow 0) = \frac{g c}{\kappa}
\label{eq:fedd_thin}
\end{equation}
as we illustrate in Figure \ref{fig:eddington_flux}.
In some works, the single-scattering Eddington flux has been derived in the context of beamed radiation from a source, as would be appropriate for a single point source in spherical symmetry, rather than isotropic plane source radiation. In the limit of beamed radiation, the appropriate single-scattering Eddington flux is
\begin{equation}
F_{\text{Edd, beamed}}(\tau) = \frac{gc}{\kappa} \frac{\tau}{1 - e^{-\tau}} = \frac{gc \langle \Sigma_{\text{gas}} \rangle}{1 - e^{-\tau}} \, .
\label{eq:fedd_beamed}
\end{equation}
If this flux is inappropriately used for an isotropic source plane of radiation, errors of order unity result. Compare the beamed single-scattering flux (blue dot-dashed) with the isotropic single-scattering flux (solid) in Figure \ref{fig:eddington_flux}.

\section{Simulation results}
\label{section:stability}
\begin{figure*}
  \includegraphics[width=\textwidth]{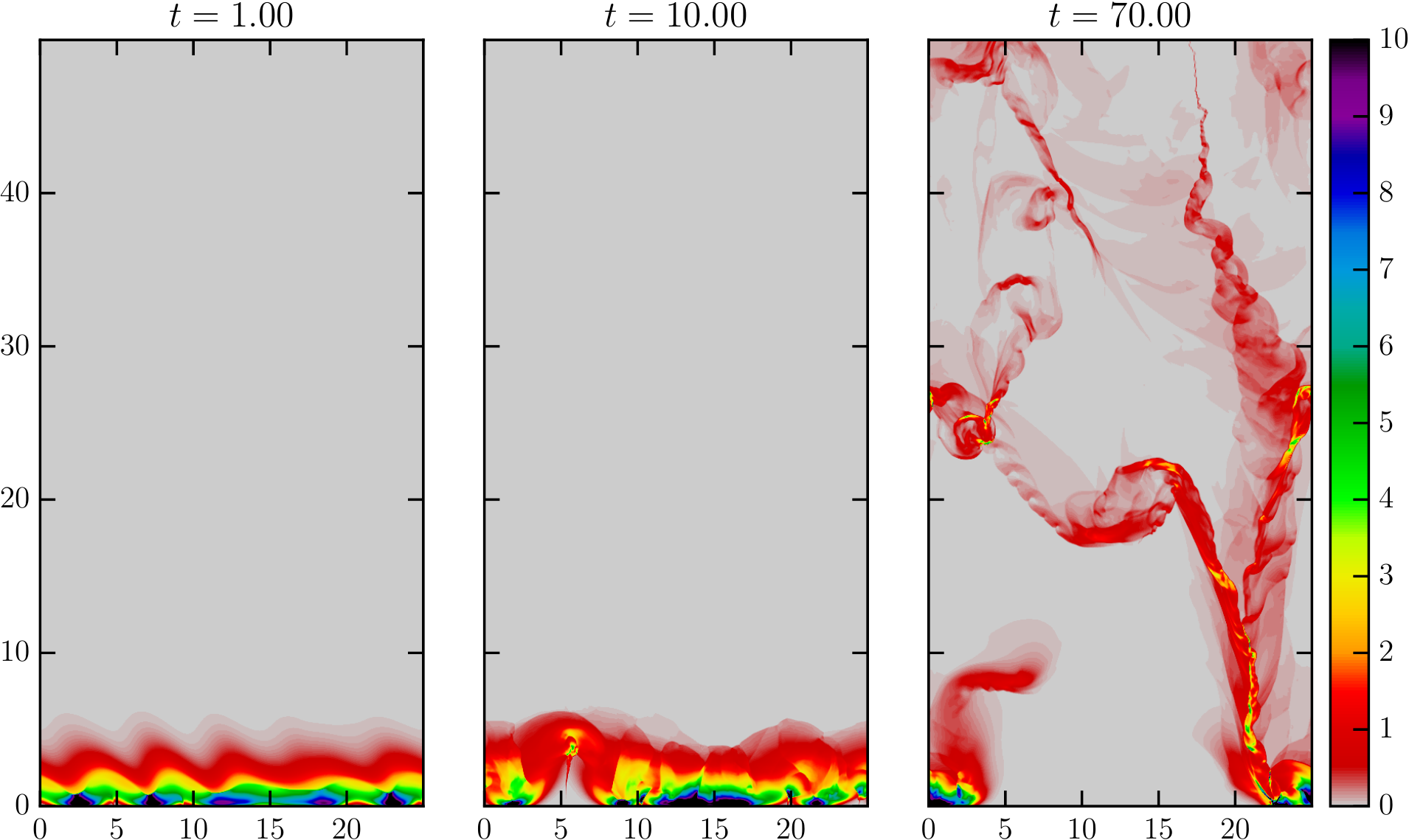}
  \caption{The density field for our fiducial simulation (simulation A) at three successive time outputs ($t=1$, $10$, $70$, respectively). The leftmost panel shows the early nonlinear evolution of the exponential atmosphere subject to the $k^{-4}$ initial forcing field.  The middle panel shows the increasing inhomogeneity in the atmosphere.  The rightmost panel shows the break-up of the atmosphere into filaments and blobs that allow the midplane flux to escape to high altitude.  The axes and colorbar scales are all in the dimensionless units defined in equations \ref{eq:dimensionless}-\ref{eq:dimensionless_erad}.}
  \label{fig:edd08_tau10_sidebyside}
\end{figure*}

\begin{figure*}
  \includegraphics[width=\textwidth]{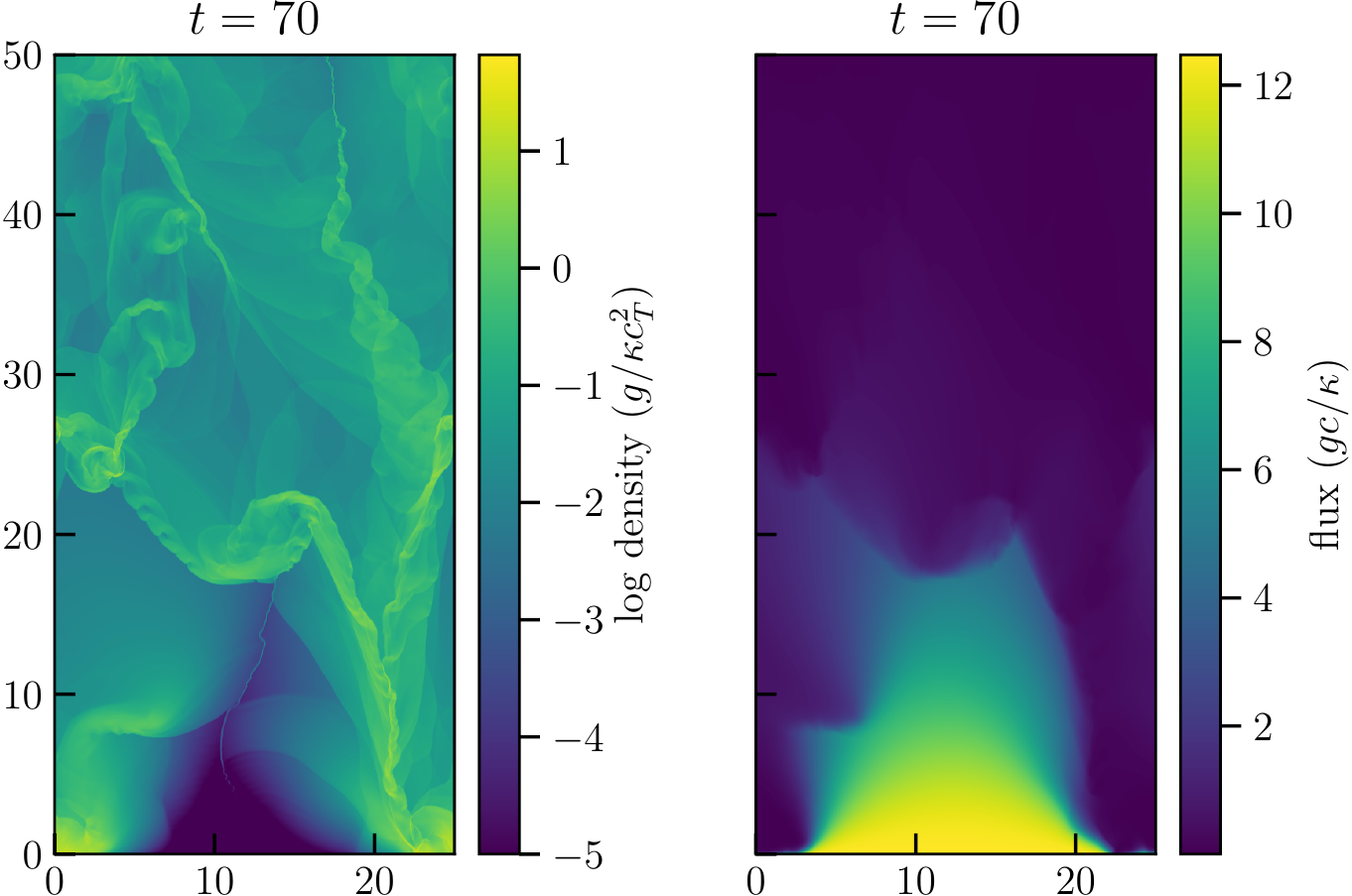}
  \caption{The log density field (compare with the right panel of Figure \ref{fig:edd08_tau10_sidebyside}) and the vertical flux (in dimensionless units; eq. \ref{eq:dimensionless_flux}) for our fiducial simulation (simulation A) at time $t=70$.}
  \label{fig:edd08_tau10_sidebyside_flux}
\end{figure*}

\begin{table*}
  \caption{Parameters used for radiation hydrodynamic simulations of the `single-scattering' limit.}
  \label{table:sims}
  \begin{tabular}{lcccccccccc}\hline
      & $\Gamma_{\text{Edd}}$ & $\tau$ & $X_{\text{max}}$ & $Y_{\text{max}}$ & $N_x$ & $N_y$ & $N_{\text{angles}}/2$ & Upper b.c. & Stochastic driving & Stable? \\
      & & & $(c_T^2 / g)$ & $(c_T^2 / g)$ & & & & & &  \\
    \hline
    \multicolumn{3}{l}{\small \textsc{Fiducial simulation}} \\
    A & 0.8 & 10 & 25 & 50 & 512 & 1024 & 1024 & Outflow & Initial ($t=0$) & Unstable \\ 
    \hline 
    \multicolumn{3}{l}{\small \textsc{Convergence tests}} \\
    A0 & 0.8 & 10 & 25 & 50 & 256 & 512 & 512 & Outflow & Initial ($t=0$) & Unstable \\ 
    A1 & 0.8 & 10 & 25 & 50 & 128 & 256 & 256 & Outflow & Initial ($t=0$) & Unstable \\ 
    A2 & 0.8 & 10 & 25 & 50 & 64 & 128 & 128 & Outflow & Initial ($t=0$) & Unstable \\ 
    A3 & 0.8 & 10 & 25 & 50 & 32 & 64 & 64 & Outflow & Initial ($t=0$) & Stable \\ 
    A4 & 0.8 & 10 & 25 & 50 & 512 & 1024 & 512 & Outflow & Initial ($t<10$) & Unstable \\ 
    A5 & 0.8 & 5 & 25 & 100 & 128 & 512 & 256 & Outflow & Initial ($t=0$) & Unstable \\ 
    A6 & 0.8 & 10 & 25 & 100 & 128 & 512 & 256 & Outflow & Initial ($t=0$) & Unstable \\ 
    A7 & 0.8 & 15 & 25 & 100 & 128 & 512 & 256 & Outflow & Initial ($t=0$) & Unstable \\ 
    B & 0.8 & 10 & 6.25 & 50 & 128 & 1024 & 512 & Outflow & Initial ($t<10$) & Unstable \\ 
    C & 0.8 & 10 & 6.25 & 50 & 128 & 1024 & 512 & Reflecting & Initial ($t<10$) & Unstable \\ 
    D & 0.8 & 10 & 4.6875 & 50 & 96 & 1024 & 512 & Outflow & Initial ($t<10$) & Unstable \\ 
    E & 0.8 & 10 & 4.6875 & 50 & 96 & 1024 & 512 & Reflecting & Initial ($t<10$) & Unstable \\ 
    F & 0.5 & 10 & 25 & 50 & 512 & 1024 & 512 & Outflow & Initial ($t<10$)& Stable \\ 
    \hline
    \multicolumn{3}{l}{\small \textsc{Parameter variations}} \\
    G & 0.9 & 0.316 & 50 & 100 & 128 & 256 & 256 & Outflow & Initial ($t<10$)& Stable \\ 
    H & 0.95 & 0.316 & 50 & 100 & 128 & 256 & 256 & Outflow & Initial ($t<10$)& Unstable \\ 
    I & 0.6 & 0.6 & 25 & 50 & 128 & 256 & 256 & Outflow & Initial ($t<10$)& Stable \\ 
    J & 0.7 & 0.6 & 25 & 50 & 128 & 256 & 256 & Outflow & Initial ($t<10$)& Unstable \\ 
    K & 0.5 & 1.0 & 25 & 50 & 128 & 256 & 256 & Outflow & Initial ($t<10$)& Stable \\ 
    L & 0.6 & 1.0 & 25 & 50 & 128 & 256 & 256 & Outflow & Initial ($t<10$)& Unstable \\ 
    M & 0.5 & 3.16 & 25 & 50 & 128 & 256 & 256 & Outflow & Initial ($t<10$)& Stable \\ 
    N & 0.6 & 3.16 & 25 & 50 & 128 & 256 & 256 & Outflow & Initial ($t<10$)& Unstable \\ 
    O & 0.7 & 3.16 & 25 & 50 & 128 & 256 & 256 & Outflow & Initial ($t<10$)& Unstable \\ 
    P & 0.8 & 3.16 & 25 & 50 & 128 & 256 & 256 & Outflow & Initial ($t<10$)& Unstable \\ 
    Q & 0.5 & 10 & 25 & 50 & 512 & 1024 & 512 & Reflecting & Initial ($t<10$)& Stable \\ 
    R & 0.6 & 10 & 25 & 50 & 512 & 1024 & 512 & Reflecting & Initial ($t<10$)& Stable \\ 
    S & 0.7 & 10 & 25 & 50 & 512 & 1024 & 512 & Reflecting & Initial ($t<10$)& Unstable \\ 
    T & 0.8 & 10 & 25 & 50 & 512 & 1024 & 512 & Reflecting & Initial ($t<10$) & Unstable \\ 
    U & 0.9 & 10 & 25 & 50 & 512 & 1024 & 512 & Reflecting & Initial ($t<10$)& Unstable \\ 
    V & 0.5 & 31.6 & 25 & 50 & 128 & 256 & 256 & Outflow & Initial ($t<10$)& Stable \\ 
    W & 0.6 & 31.6 & 25 & 50 & 128 & 256 & 256 & Outflow & Initial ($t<10$)& Stable \\ 
    X & 0.7 & 31.6 & 25 & 50 & 128 & 256 & 256 & Outflow & Initial ($t<10$)& Stable \\ 
    Y & 0.8 & 31.6 & 25 & 50 & 128 & 256 & 256 & Outflow & Initial ($t<10$)& Unstable \\ 
    \hline
  \end{tabular}
\end{table*}

The parameters for the simulations are given in Table \ref{table:sims}. We choose box dimensions that are an order of magnitude larger than the (gravitational) thermal pressure scale height $y_0 = c_T^2/g$ in order to resolve the long-wavelength instabilities suggested by our perturbative analysis (section \ref{section:perturbation_theory}). We choose optical depths of $\tau = 10^{-0.5}-10^{1.5}$ that logarithmically sample the lower range of expected optical depths inferred from observations of star-forming galaxies (see section \ref{section:discussion}) and spans the optically-thin to optically-thick regimes.  We use a number of angles for the radiation that is twice the number of zones in the vertical direction, ensuring that the angular distribution of the specific intensity is well-resolved in the optically-thin limit (Appendix \ref{appendix:resolution}).

We choose initial conditions of gravitational hydrostatic equilibrium (when the thermal pressure gradient balances gravity, i.e., an exponential gas pressure vertical profile) and then turn on solenoidal stochastic turbulent driving with a time-dependent external forcing $\vec{f}$.  At each timestep when it is turned on, the external forcing $\vec{f}$ is computed as an independent realization of a Gaussian random field with $f(k) \propto k^{-4}$ and no compressive modes (i.e., $\div{\vec{f}} = 0$).\footnote{We use the implementation included in \textsc{Athena} in the file \texttt{turb.c}.}

\subsection{Fiducial simulation}
\label{subsection:fiducial_sim}
For our fiducial simulation (simulation A), we apply a strong impulsive perturbation at the initial timestep, and we find the ensuing unstable behavior drives transonic turbulent velocities driven by radiation pressure \emph{without} further driving from the forcing field. In the leftmost panel of Figure \ref{fig:edd08_tau10_sidebyside}, which shows the full vertical and horizonal extent of the simulation, we see the imprint of the initial perturbations at time $t=1$ (left), instabilities altering the nonlinear evolution by time $t=10$ (middle), causing the density profile to break up into high- and low-column-density regions, and at $t=70$ (right) the initial density profile is fully broken up into jets and filamentary features with transonic or mildly supersonic velocities that are characteristic of the two-dimensional unstable behavior of our idealized system. The breakup of the density profile into filaments and blobs allows for the escape of substantial amounts of radiation to altitudes that are very high compared to the gas pressure scale height.

We see this relationship between flux and density more clearly in Figure \ref{fig:edd08_tau10_sidebyside_flux}, where the left panel shows the same density field as in the rightmost panel of Figure \ref{fig:edd08_tau10_sidebyside} but logarithmically scaled, and the right panel shows the vertical flux for the same simulation time. The low density region in the lower middle of the simulation box (dark blue in the log density field) enables the flux to escape to a high height (tens of $c_T^2 / g$ scale heights) and levitates the filaments that absorb the majority of the flux at the edges of this low-density region. This relationship between flux and density is characteristic of the late-time highly-nonlinear behavior of this simulation, as well as all other simulations we have run with a strong initial impulsive perturbation that show evidence of instabilities.

\begin{figure}
  \includegraphics[width=\columnwidth]{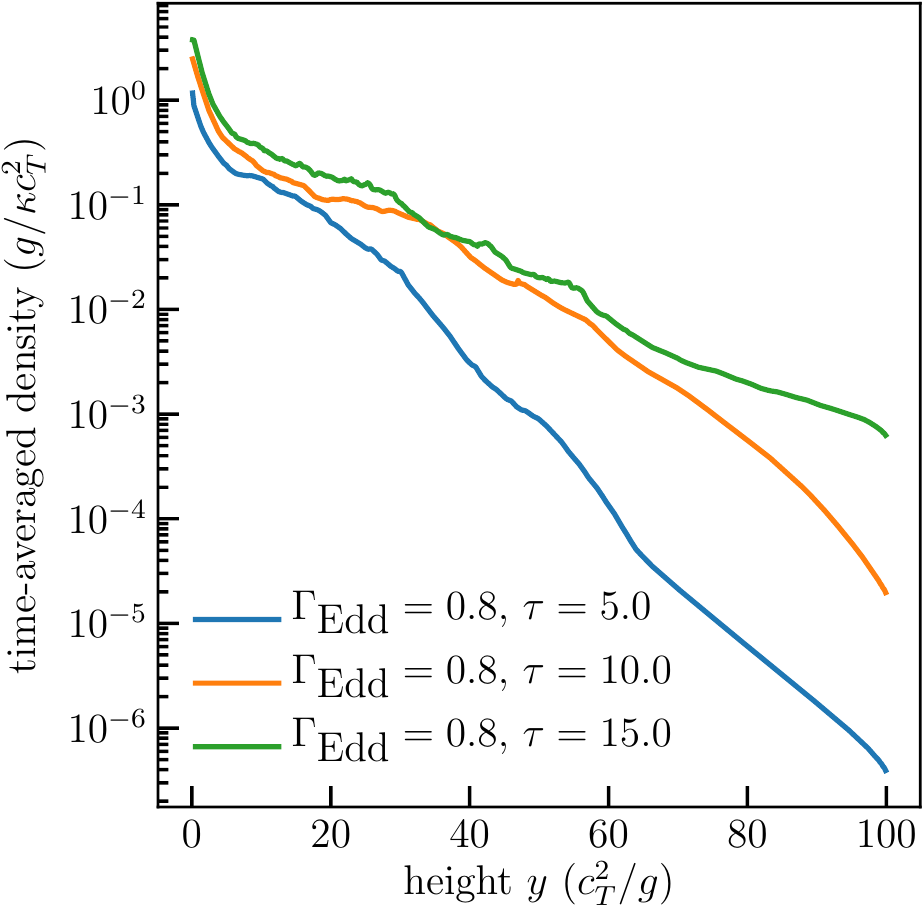}
  \caption{The vertical density profile for simulations A5, A6, and A7 in Table \ref{table:sims}, averaged over times $50 \, c_T/g < t < 300 \, c_T/g$.}
  \label{fig:density_profiles}
\end{figure}
In Figure \ref{fig:density_profiles}, we show the time-averaged vertical density profile for a lower-resolution simulation otherwise equivalent to the fiducial simulation, except with twice the vertical extent, which enables us to resolve the characteristic scale height associated with the turbulent dynamical equilibrium for this particular set of parameters in the unstable regime. We also show the density profiles for lower and higher optical depth simulations, illustrating that the density profile scale height and normalization increases with optical depth. We discuss these additional simulations in section \ref{subsection:variations} and we discuss the character of the dynamical equilibrium in section \ref{subsection:summary}.

We note that the physical resolution of our fiducial simulation is
\begin{align}
\Delta x = 0.04 \, \text{pc} \left( \frac{T}{300 \,K} \right) \left( \frac{\Sigma}{100 \, M_{\sun} \text{pc}^{-2}} \right)^{-1} \, .
\end{align}
We discuss the resolution dependence of our simulations in Appendix \ref{appendix:resolution}.

\subsection{Variations of optical depth and Eddington ratio}
\label{subsection:variations}
By running variants of this fiducial simulation (the simulations in the second section of Table \ref{table:sims}), with varying Eddington ratio $\Gamma_{\text{Edd}}$ and optical depth $\tau$, we find that there is a critical Eddington ratio $\Gamma_{\text{Edd, crit}}(\tau)$ \emph{below} which large perturbations decay and the long-term evolution relaxes to hydrostatic equilibrium and \emph{above} which we obtain self-sustaining transonic or mildly supersonic turbulence driven solely by radiation pressure (i.e., the behavior seen in the rightmost panel of Figure \ref{fig:edd08_tau10_sidebyside}).  For these simulations, we apply the forcing at each timestep until $t = 10 \, (c_T/g)$ without radiation, when the system's velocity dispersion approaches a statistical steady-state, which we take to indicate an approximate dynamical equilibrium.  We then disable the stochastic driving for the remainder of each simulation, and turn on a source of radiation at the lower boundary with a flux given by equation \ref{eq:eddington_ratio} to obtain the desired Eddington ratio.

\begin{figure}
  \includegraphics[width=\columnwidth]{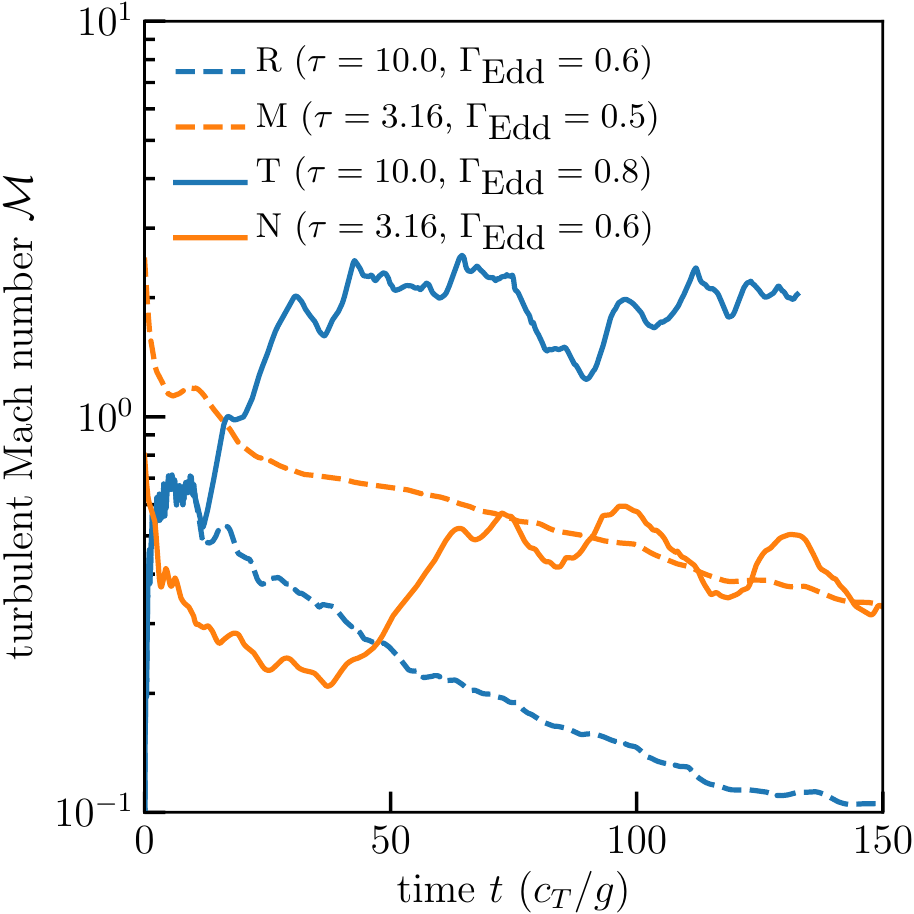}
  \caption{The turbulent Mach number $\mathcal{M} = \delta v / c_T$ (see eq. \ref{eq:dv}) as a function for time for a subset of our simulations (simulations R, M, T, and N listed in Table \ref{table:sims}). The solid lines show $\mathcal{M}$ for the unstable simulations, which produce self-sustaining turbulent velocities through radiatively-driven instabilities.  The dashed lines show the velocity dispersion for the stable simulations, for which the turbulent velocities decay after the external stochastic driving is turned off.}
  \label{fig:turb_dv_all}
\end{figure}

To illustrate the behavior of simulations above and below the critical Eddington ratio, we show the turbulent velocity dispersion as a function of time for two pairs of simulations that lie on either side of the boundary between  self-sustaining turbulence and decaying turbulence.  Two simulations with self-sustaining turbulence are shown as solid lines and two simulations with the same optical depth but with a lower Eddington ratio that exhibit decaying turbulence are plotted as dotted lines in Figure \ref{fig:turb_dv_all}.  We define the (mass-weighted, 1D-equivalent) turbulent velocity dispersion $\delta v$ as
\begin{equation}
\left( \delta v \right)^2 = \frac{1}{M} \int \frac{1}{2} \left( (v_x - \langle v_x \rangle)^2 + (v_y - \langle v_y \rangle)^2 \right) \rho \, dV \, , \label{eq:dv}
\end{equation}
where $M$ is the total mass inside the simulation box, $v_x$ and $v_y$ are the velocity components, $\rho$ is the mass density, $dV$ is the 2D volume element, and $\langle v_i \rangle = M^{-1} \int \rho v_i \, dV$. We see that for $\Gamma_{\text{Edd}}=0.8$ and $\tau=10$, the turbulent velocity dispersion appears to asymptote to $\sim 2 c_T$ over many tens of $c_T/g$ timescales.

\begin{figure}
  \includegraphics[width=\columnwidth]{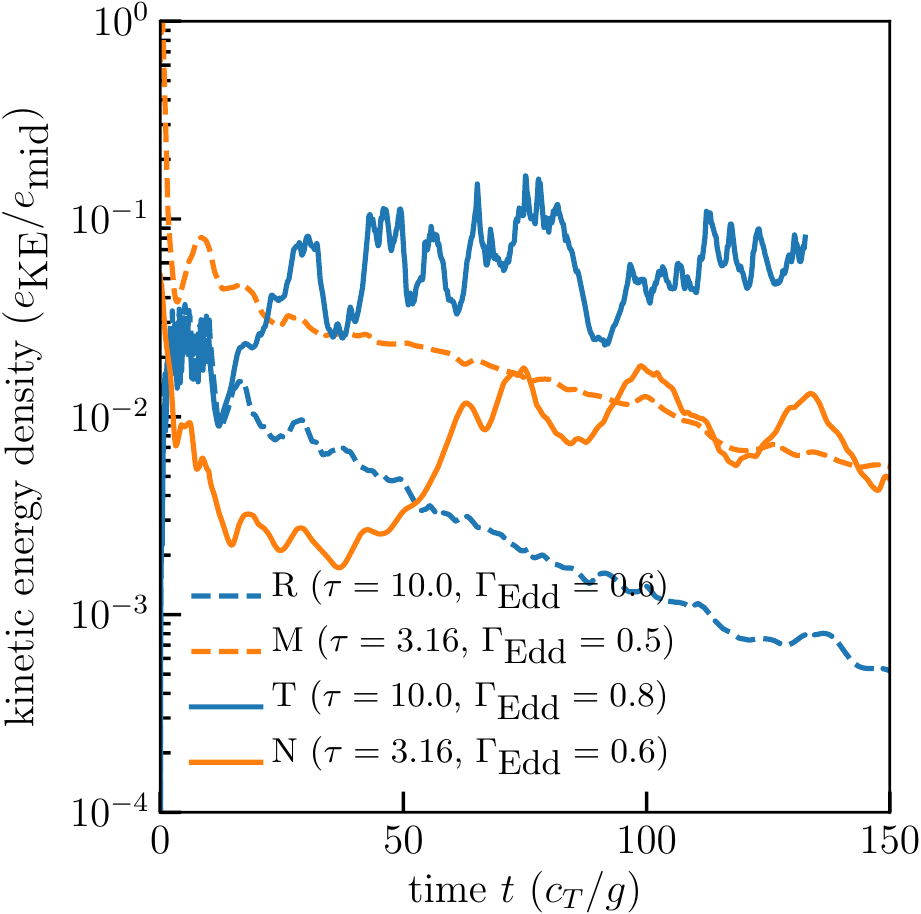}
  \caption{The turbulent energy density (see eq. \ref{eq:erad}) in units of the midplane radiation energy density (eq. \ref{eq:emid}) as a function of time for a subset of the simulations (simulations R, M, T, and N listed in Table \ref{table:sims}). The solid lines show the simulations for which turbulence is self-sustained, and the dotted lines show the simulations for which turbulence decays.}
  \label{fig:erad_time}
\end{figure}

The turbulent energy density of the same simulations as a function of time is shown in Figure \ref{fig:erad_time}. The mass-weighted turbulent energy density
\begin{align}
\delta e_{\text{rad}} = \frac{1}{M} \int \left[ \frac{1}{2} \rho \left( (v_x - \langle v_x \rangle)^2 + (v_y - \langle v_y \rangle)^2 \right) \right] \rho \, dV \label{eq:erad}
\end{align}
is an order of magnitude less than the radiation energy density at the lower boundary
\begin{align}
e_{\text{mid}} = (2\pi / c) \, \Gamma_{\text{Edd}} F_{\text{Edd}}(\tau) \, . \label{eq:emid}
\end{align}
Thus, the turbulent energy density of our idealized disk is \emph{not} in equipartition with the radiation energy density at the midplane of the disk, which drives the motion. This energy balance discrepancy may be due to much of the radiation flux acting to levitate the disk, which, in hydrostatic equilibrium, does no work on the fluid. We have verified that, in a time-averaged sense, very little of the flux escapes from the atmosphere, consistent with our interpretation of the time-averaged vertical density profiles in section \ref{subsection:summary}.

\begin{figure}
  \includegraphics[width=\columnwidth]{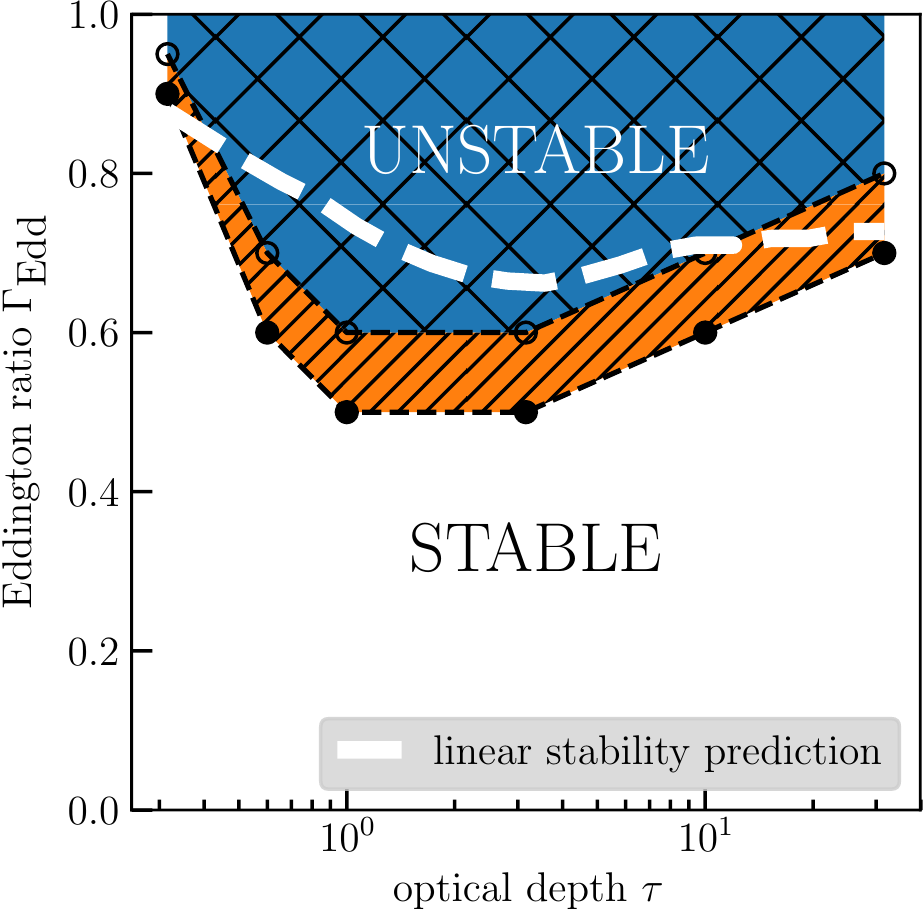}
  \caption{The stability curve in the optical depth-Eddington ratio parameter space, as determined empirically from our simulations that assume isotropic midplane radiation (listed in Table \ref{table:sims}). The white dotted line shows the linear stability prediction for beamed radiation derived in section \ref{section:perturbation_theory}, as previously shown in Figure \ref{fig:perturbative_stability}.}
  \label{fig:stability}
\end{figure}

We show the stable (lower white) and unstable (upper cross-hatched blue) regions of parameter space empirically determined by our simulations in Figure \ref{fig:stability}.  The open circles indicate simulations with parameter values that are found to self-sustain turbulence after the initial perturbation, and those with filled circles are those for which the perturbations decay and produce a stable hydrostatic atmosphere in the presence of radiation forces (even though they may have inversions in their vertical density profiles).  The indeterminate region (in which we did not run any simulations, as explained below) is the shaded orange region. For $\tau=0.316$, we find the transition to instability is between $\Gamma_{\text{Edd}} = 0.9-0.95$. For $\tau=0.6$, we find it is between $\Gamma_{\text{Edd}}=0.6-0.7$. For $\tau=1$, we find it is between $\Gamma_{\text{Edd}}=0.5-0.6$. For $\tau=3.16$, the limit is between $\Gamma_{\text{Edd}} = 0.5-0.6$.  For $\tau=10$, this limit is between $\Gamma_{\text{Edd}} = 0.6-0.7$.  For $\tau=31.6$, this limit is between $\Gamma_{\text{Edd}} = 0.7-0.8$. This limit appears to be qualitatively consistent with the stability curve in the $(\Gamma_{\text{Edd}}, \tau)$ parameter space that we obtained by perturbative analysis for beamed radiation, shown as the dashed white line in Figure \ref{fig:stability} (section \ref{section:perturbation_theory}). 

We have tested, through many additional simulations not otherwise described in this paper, that there do not appear to exist any `islands' of stability (instability) within the unstable region (stable region) of parameter space.  We do not systematically carry out simulations within the indeterminate region of parameter space (orange, single-hatched region of Figure \ref{fig:stability}) because the transition from stability to instability is not sharp for simulations of finite box size and finite time, as indicated by test simulations with parameters near this transition. Given the uncertainties in computing the Eddington ratios of galaxies and the idealized relationship between our models and real galaxies, a more precise estimate of the instability region in this parameter space is not particularly useful.


\subsection{Summary of results}
\label{subsection:summary}
\begin{figure}
  \includegraphics[width=\columnwidth]{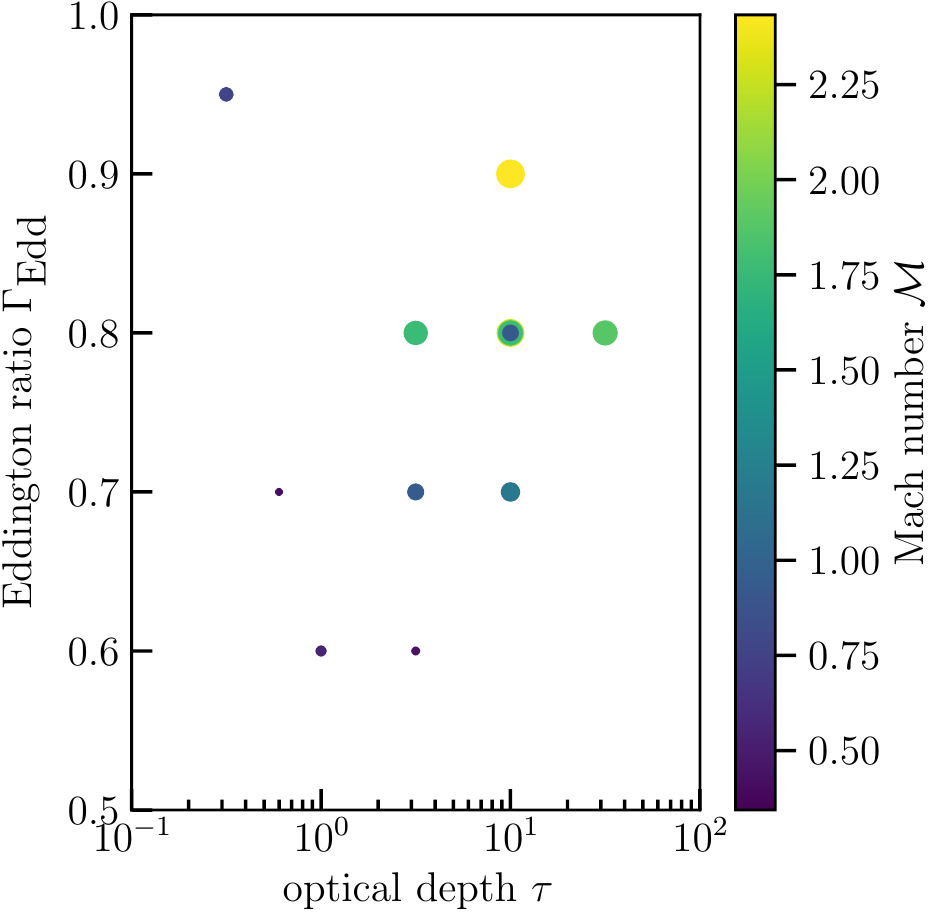}
  \caption{The time-averaged ($t > 50 \, c_T/g$) velocity dispersion as a function of Eddington ratio $\Gamma_{\text{Edd}}$ and optical depth $\tau$ for the simulations in Table \ref{table:sims} where the driving is for time $t < 10$. The size of the circles scales with the Mach number of the simulations.}
  \label{fig:summary_dv}
\end{figure}

\begin{figure}
  \includegraphics[width=\columnwidth]{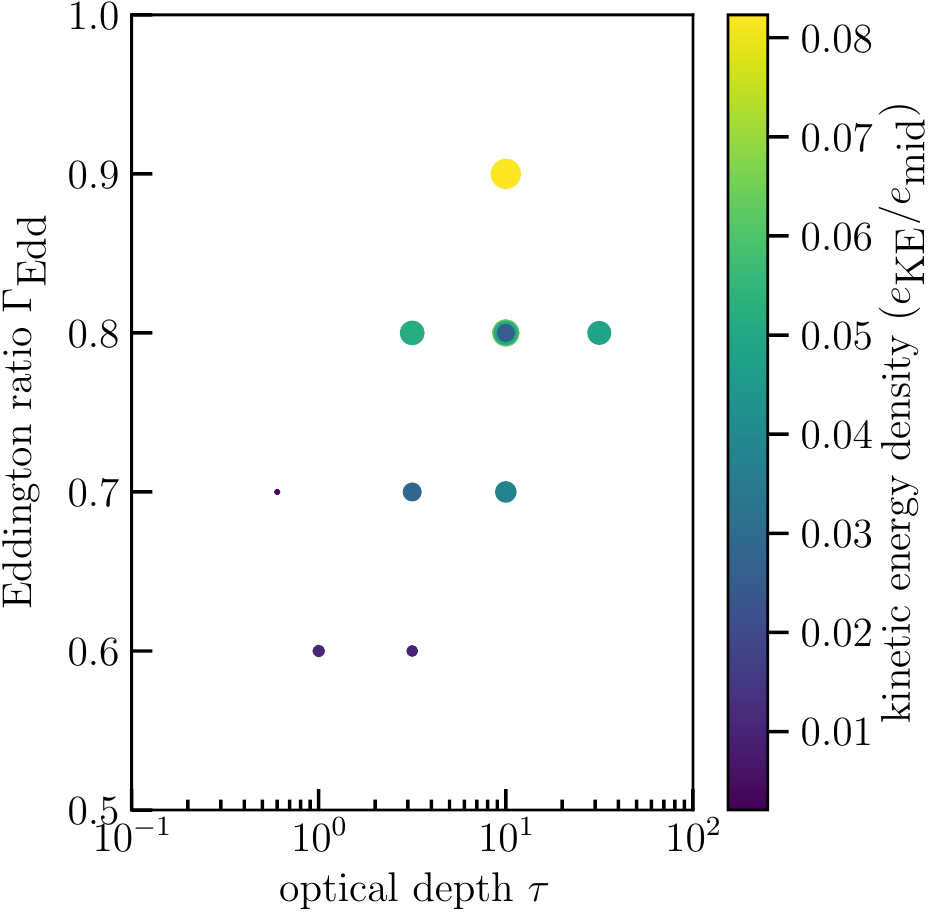}
  \caption{The time-averaged ($t > 50 \, c_T/g$) turbulent energy density (eq. \ref{eq:erad}) in units of the midplane radiation energy density (eq. \ref{eq:emid}) as a function of Eddington ratio $\Gamma_{\text{Edd}}$ and optical depth $\tau$ for the simulations in Table \ref{table:sims} where the driving is for time $t < 10$. The size of the circles scales with the kinetic energy density of the simulations.}
  \label{fig:summary_ke}
\end{figure}

We summarize the time-average velocity dispersion and the time-average turbulent energy density of all of our simulations that we run with forcing turned on at times $t < 10$ in Figures \ref{fig:summary_dv} and \ref{fig:summary_ke}. Generally, we find that the velocity dispersion and turbulent energy density increase with both optical depth and Eddington ratio. For example, at $\tau=10$ with the fiducial box size, we find that for $\Gamma_{\text{Edd}}=0.7$, $\delta v \sim 1.4$ (averaged over $50 < t < 100$); for  $\Gamma_{\text{Edd}}=0.8$, we find  $\delta v \sim 1.9$; and for $\Gamma_{\text{Edd}}=0.9$, we find $\delta v \sim 2.4$.  Likewise, at $\Gamma_{\text{Edd}} = 0.7$, we find that for $\tau = 0.6$, $\delta v \sim 0.5$; for $\tau = 3.16$, $\delta v \sim 0.9$; and for $\tau = 10$, $\delta v \sim 1.4$ (eq. \ref{eq:dv}). In all parameter regimes simulated, we find that $\frac{1}{2} \rho \, \delta v^2 \ll e_{\text{rad}} = 2 \pi (F_{\text{mid}}/c)$, in contrast to the \emph{ansatz} that turbulence will reach energy equipartition with radiation when $F_{\text{mid}} \sim F_{\text{Edd}}$ (i.e., $\frac{1}{2} \rho \, \delta v^2 \sim e_{\text{rad}}$; see eqs. 4, 9, and 10 in \citealt{Thompson_2005} for a Toomre $Q=1$ disk). We find that the velocity dispersion $\delta v$ depends on the incident flux $F$ as $\delta v \sim \sqrt{F}$, saturating at $\sim 2 \, c_T$ for the highest midplane fluxes in the parameter space that we simulate. We speculate that this energy balance \emph{ansatz} fails because the radiation does no work when levitating the fluid in hydrostatic equilibrium. If this explanation is correct, it suggests that the mean properties of the simulations may be largely explained by a model of hydrostatic equilibrium in the presence of radiation and gravity, with a subdominant turbulent pressure contribution (see section \ref{subsection:summary} on time-averaged density profiles).

This lack of equipartition can also be explained by noting that the flux cannot be arbitrarily high without driving a strong wind that will inevitably expel all of the gas. This is derived as a maximum limit on $F_{\text{mid}}$ for a hydrostatic atmosphere in section \ref{section:perturbation_theory}. Intuitively, this is the case even in the presence of turbulence (whether self-generated or not) because the single-scattering Eddington flux is always greater than the optically thin Eddington flux $g c / \kappa$ (eq. \ref{eq:fedd_ss}; see also Figure \ref{fig:eddington_flux}), and therefore there can exist systems (for Eddington ratios near unity) where the incident flux is super-Eddington with respect to the optically-thin Eddington flux $gc / \kappa$ but sub-Eddington with respect to the single-scattering Eddington flux $F_{\text{Edd}}$ (i.e., when $\Gamma_{\text{Edd}} F_{\text{Edd}} > gc / \kappa$). In such a regime (applicable to simulation A), any very dense columns that are temporarily sub-Eddington (due to turbulent density fluctuations that increase their density such that they self-shield) will be dispersed into super-Eddington columns as they sink, spread out horizontally, and become optically thin, as observed in our simulations.

We can estimate the expected turbulent Mach number at order-of-magnitude for the optically-thick single-scattering Eddington limit with basic dimensional considerations. Equating the rate of shock dissipation with the rate of work done by radiation on the fluid
\begin{align}
\left\langle \dot e_{\text{turb}} \right\rangle = \frac{de_{\text{out}}}{dt} &= \frac{de_{\text{in}}}{dt} = \langle \v{v} \cdot \frac{\kappa \rho}{c} \v{F}_{\text{rad}} \rangle \label{eq:shock_diss} \, ,
\end{align}
where the angle brackets indicate the expectation value over space or time, and dots indicate partial derivatives with respect to time. We approximate the expectation values by replacing the state variables with characteristic scales (i.e., replacing $\rho$ with $\rho_0$), assuming that $\v{v} \sim \delta v$, taking the radiation flux to be the optically-thick isotropic Eddington flux at the midplane for a system with optical depth $\tau$ and Eddington ratio $\Gamma_{\text{Edd}}$ (eq. \ref{eq:fedd_thick}), and, extending the dimensional analysis presented in \cite{Gammie_1996}, introduce the dissipation length scale $L$:
\begin{align}
\frac{1}{2} \rho_0 \, \delta v^2 \left( \frac{L}{\delta v} \right)^{-1} = \delta v \frac{\kappa \rho_0}{c} \Gamma_{\text{Edd}} \frac{3}{2} \frac{g c}{\kappa} \tau \, .
\end{align}
We replace the characteristic scales with those defined in section \ref{section:methods}, and obtain an expression for the turbulent velocity dispersion $\delta v$:
\begin{align}
\delta v^2 / c_T^2 &= 3 \Gamma_{\text{Edd}} \tau \\
\mathcal{M} &\sim 1.7 \sqrt{\Gamma_{\text{Edd}} \tau} \label{eq:edot_mach_prediction}
\end{align}
where we have further assumed that the dissipation length scale $L$ is equal to the gas pressure scale height $c_T^2 / g$. This equation is only sensible if $\Gamma_{\text{Edd}} \le 1$, because otherwise the entire column of fluid will be launched in a wind. Thus, the turbulent Mach number is constrained by physical and dimensional considerations to be $\lesssim \sqrt{\tau}$ for $\tau \gg 1$.

We now compare to the Mach number expected under the assumption that the turbulent kinetic energy density is in equipartion with the midplane radiation energy density:
\begin{align}
e_{\text{turb}} &\sim e_{\text{rad}}\, .
\end{align}
We parameterize the flux in terms of the optically-thick Eddington flux at the midplane to obtain
\begin{align}
\frac{1}{2} \rho_0 \delta v^2 &\sim \frac{2 \pi}{c} \Gamma_{\text{Edd}} \frac{3}{2} \frac{g c}{\kappa} \tau
\end{align}
and we again substitute the characteristic scales defined in section \ref{section:methods} to obtain
\begin{align}
\delta v^2 / c_T^2 &\sim 6 \pi \Gamma_{\text{Edd}} \tau \\
\mathcal{M} &\sim 4.3 \sqrt{\Gamma_{\text{Edd}} \tau} \, . \label{eq:e_mach_prediction}
\end{align}
This assumption of energy equipartition thus produces a turbulent kinetic energy density (since $e_{\text{turb}} \propto \mathcal{M}^2$) greater by a factor of $2\pi$ compared to equating the expectation values of the relevant terms in the momentum equation. As is apparent from the overall Mach number normalization in Figure \ref{fig:summary_dv}, our simulations agree better with the prediction obtained by equating the time derivatives of the energy densities (eq. \ref{eq:edot_mach_prediction}), rather than the energy densities themselves (eq. \ref{eq:e_mach_prediction}). However, we note that the scaling with Eddington ratio and optical depth is identical in both formulations.

\begin{figure}
  \includegraphics[width=\columnwidth]{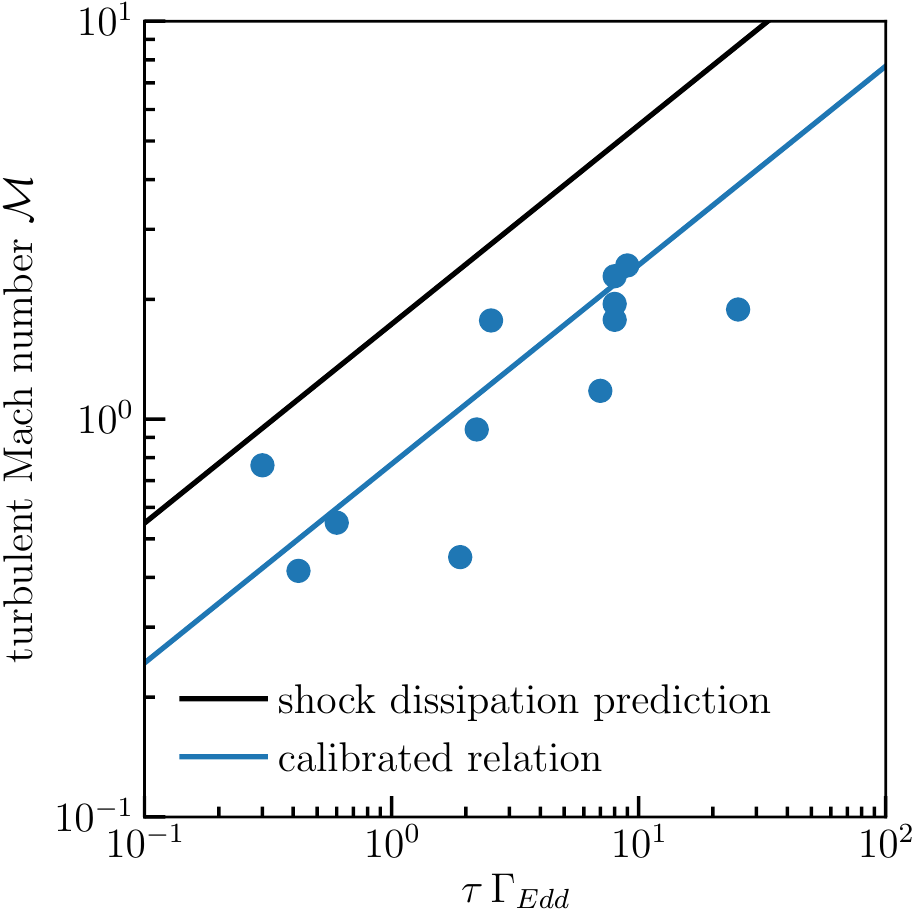}
  \caption{The $\tau \Gamma_{\text{Edd}}$-$\mathcal{M}$ scaling relation, as predicted by eq. \ref{eq:edot_mach_prediction} (solid black line) and as rescaled to our simulations (solid blue line). The time-average Mach numbers from our individual simulations are shown by the blue points. Our simulations suggest a turbulent dissipation scale $\sim 5$ times smaller than we assumed in \ref{eq:edot_mach_prediction}.}
  \label{fig:tau_mach_scaling}
\end{figure}
Since the prediction from our first argument (eq. \ref{eq:edot_mach_prediction}) is found to agree at the factor-of-few level with our simulations and is based on a reasonable description of the physics involved, we use our simulations to rescale our order-of-magnitude predictions (Figure \ref{fig:tau_mach_scaling}). Since our simulated squared turbulent Mach numbers are about a factor $\sim 5$ smaller than predicted based on the dimensional analysis argument equating time derivatives of energy densities (which, taking the model at face value, implies a dissipation length scale $L$ that is $\sim 5$ times \emph{smaller} than the assumed $L = c_T^2/g$), we predict an increase in the squared turbulent Mach number $\mathcal{M}^2$ in systems where the optically-thick single-scattering limit applies and the system is within the unstable parameter regime of Figure \ref{fig:stability} by the additive factor
\begin{align}
\Delta \mathcal{M}^2 &\approx 3 \, \Gamma_{\text{Edd}} \tau \, / \, 5 = 0.6 \, \Gamma_{\text{Edd}} \tau \, .
\end{align}
Assuming that the turbulence produced by radiation pressure behaves as a linearly-additive energy source in the sense of eq. \ref{eq:shock_diss}, the rescaling is appropriately done in the square of the Mach number since $\mathcal{M}^2$ is proportional to the turbulent energy density. This calibration predicts that for the extreme limit of astrophysically reasonable optical depths and Eddington ratios ($\Gamma_{\text{Edd}} \tau \, \sim 10^3$), radiation pressure may produce $\mathcal{M} \sim 24$ turbulence in a $\tau \sim 10^3$ and $\Gamma_{\text{Edd}} \sim 1$ system. However, we caution that this prediction is an extrapolation from the lower-optical-depth regime in which we have conducted simulations and we cannot rule out that the calibration factor itself may be a function of optical depth, and more importantly, that such predictions only apply for systems within the unstable parameter regime identified in Figure \ref{fig:stability}.

Our arguments above about shock dissipation may explain the vertical density profiles shown in Figure \ref{fig:density_profiles}. If we assume that the scale height of the time-average density profile $h_{\text{turb}}$ is given at order-of-magnitude by $\mathcal{M}^2 c_T^2 / g (1 - \Gamma_{\text{Edd}})$ instead of $c_T^2/g$ (i.e., assuming that the turbulent velocities act as a pressure in the same manner as gas pressure and that the effective gravity is reduced to obtain the appropriate mass-weighted radiative acceleration), then we have
\begin{align}
h_{\text{turb}} &\sim \frac{\mathcal{M}^2}{1-\Gamma_{\text{Edd}}} \frac{c_T^2}{g} \sim \frac{\Gamma_{\text{Edd}} \tau}{1 - \Gamma_{\text{Edd}}} \frac{c_T^2}{g} \, , \label{eq:h_turb}
\end{align}
predicting that for $\tau = 10$ and $\Gamma_{\text{Edd}} = 0.8$, the turbulent scale height is $\sim 40 \, c_T^2/g$. We see that this agrees reasonably well with the scale height inferred from the vertical density profile for these parameters shown in Figure \ref{fig:density_profiles}. Further, this \emph{ansatz} predicts that the turbulent scale height is proportional to optical depth, which is qualitatively supported by the trend with optical depth also shown in Figure \ref{fig:density_profiles}. In the presence of radiation pressure, the scale height is substantially inflated compared to what is expected from the turbulent pressure alone, due to an effective gravity that is smaller than $g$. This fact should be taken into account in interpreting observations of systems that lie within the unstable parameter space we identify in Figure \ref{fig:stability}.

\section{Discussion}
\label{section:discussion}

\subsection{Application to star-forming disks}
\begin{figure}
  \includegraphics[width=\columnwidth]{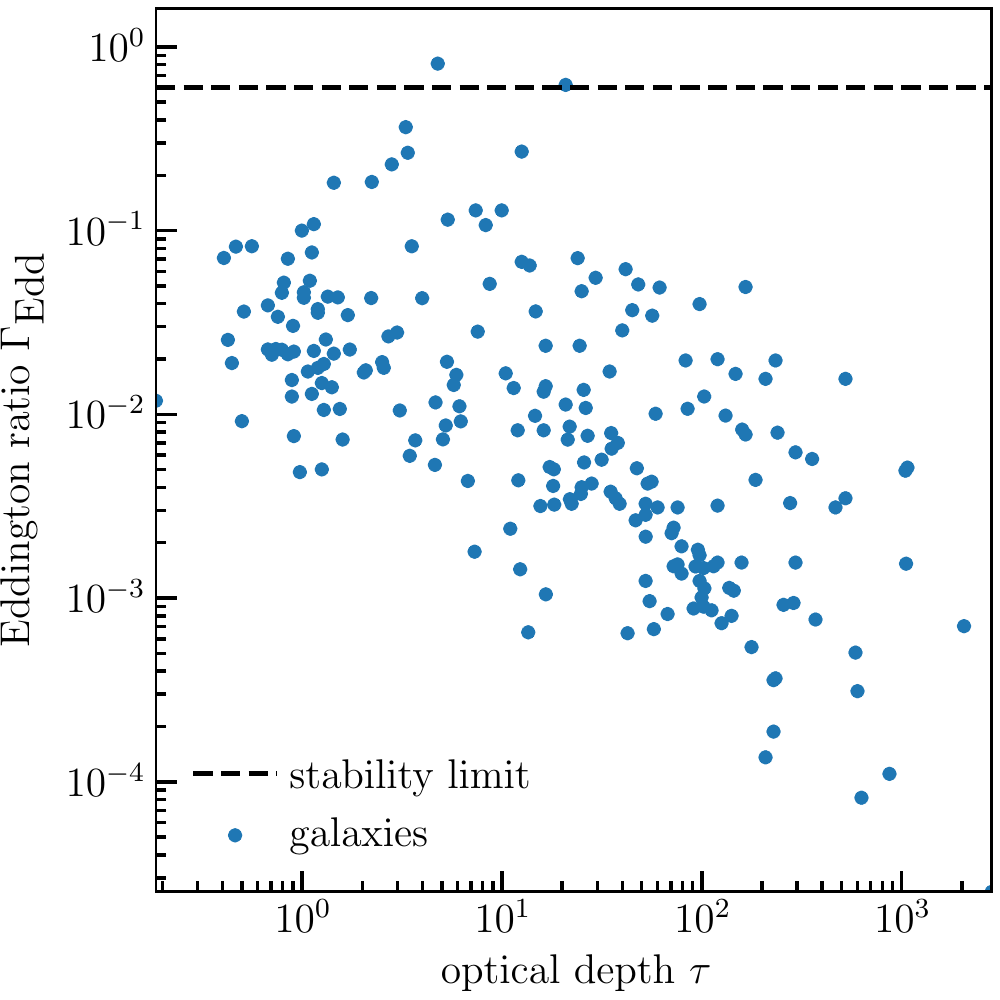}
  \caption{The Eddington ratio and optical depth as inferred from the sample of galaxies compiled in \citealt{Krumholz_2014}, assuming a gas mass fraction $f_g=0.3$ (eqs. \ref{eq:obs_tau}-\ref{eq:obs_tir}). The horizontal dashed line is an approximate fit to the minimum of the stability curve shown in Figure \ref{fig:stability}.}
  \label{fig:bimodal_aco_galaxies}
\end{figure}
In Figure \ref{fig:bimodal_aco_galaxies}, we examine the galaxy-averaged distribution of Eddington ratios and optical depths for the sample of galaxies compiled in \cite{Krumholz_2014} (based on observations from \citealt{Kennicutt_1998,Bouche_2007,Daddi_2008,Daddi_2010,Genzel_2010,Tacconi_2013,Davis_2014a}). We convert the observed star formation rate and gas surface densities into these quantities using
\begin{align}
\tau &= \kappa \Sigma_g / 2 \, , \label{eq:obs_tau}\\
F_{\text{Edd}} &= 2 \pi G c f_g^{-1} \, \Sigma_g^2 \\
&\times \frac{3}{2} \left[ 1 + e^{-\tau} \left( \frac{\tau}{2} - \frac{\tau^{2}}{2} - 1 \right) + \frac{\tau^3}{2} \int_{\tau}^{\infty} e^{-t}/t \, dt \right]^{-1} \label{eq:fedd_ss_galaxy} \, ,
\end{align}
and
\begin{align}
\Gamma_{\text{Edd}} &= F_{\text{TIR}} / F_{\text{Edd}} \, , \label{eq:obs_tir}
\end{align}
where we have assumed a gas mass fraction $f_g = 0.3$, a flux-mean opacity $\kappa = 10^3$ cm$^2$ g$^{-1}$, and $F_{\text{TIR}}$ is the total observed IR flux.

Figure \ref{fig:bimodal_aco_galaxies} shows the inferred Eddington ratios and optical depths for the sample and the boundary between stability and instability we have identified in this paper. Subject to uncertainties in dust-to-gas ratio, $\alpha_{\text{CO}}$ conversion factor, and gas-to-total-mass ratio $f_g$, we find that only two galaxies in the sample are possibly above this limit (using the bimodal $\alpha_{\text{CO}}$ conversion factor used for a reanalysis of the same dataset by \citealt{TK_2016}).  The vast majority of galaxies are highly sub-Eddington on average and are therefore not in the region of self-sustaining turbulence driven by radiation pressure identified in Figure \ref{fig:stability}.  However, we note that the observations used in Figure \ref{fig:bimodal_aco_galaxies} do not resolve the galaxies. Sub-regions of galaxies, especially central star-forming regions, may be much closer to Eddington, as suggested by comparing the local interstellar radiation field ($\sim 1 \, G_0$) with that inferred for the Galactic center ($\sim 10^3 \, G_0$; see \citealt{Lis_2001}). Resolved observations of star-forming galaxies should be able to test this scenario, especially when combined with dust modeling of IR observations to infer the radiation energy density.  Previously, several authors have considered the dynamical effects of radiation pressure in giant molecular clouds both theoretically and observationally \citep{Scoville_2001,MQT_2010,Lopez_2011,Lopez_2014,Raskutti_2016} and found the single-scattering Eddington ratio may approach or exceed unity during the cloud's dynamical evolution, suggesting that resolved extragalactic observations may also show star-forming sub-regions to be at or above Eddington.

\subsection{Application to galactic winds}
For simulations where we resolve the turbulent scale height (eq. \ref{eq:h_turb}; see also Figure \ref{fig:density_profiles}), we find no physically significant mass loss (we do find a change in mass of order one part in $10^3$ over $300 \, c_T/g$ timescales, but this is essentially insignificant and may also be due in part to numerical effects caused by our boundary conditions). This suggests that winds are only driven by radiation pressure in the single-scattering limit when the mean mass-weighted acceleration is greater than the gravitational acceleration (i.e., $\Gamma_{\text{Edd}} > 1$), or possibly when the turbulent scale height (inflated relative to the gas pressure scale height) approaches the disk radius, thus becoming large enough that the plane-parallel approximation breaks down. We note that simulations with box height $Y_{\text{max}}$ less than the turbulent scale height (e.g., simulation A in Table \ref{table:sims}), we find that radiative acceleration does drive mass loss from the box. 

\cite{TK_2016} propose that globally sub-Eddington systems may still drive winds due to \emph{local} variations in the mass-weighted radiative acceleration, thus modifying the global dynamics due to such local variations of $\Gamma_{\text{Edd}}$ that should occur with a wide distribution of column densities produced by highly supersonic turbulence.  That is, low column density sightlines in a turbulent medium may be super-Eddington even though the system is sub-Eddington on average, due to the linear scaling of $F_{\text{Edd}}$ on $\Sigma_{\text{gas}}$ in the single-scattering limit (eq. \ref{eq:fedd_ss}).

Since we do not observe such an effect in our simulations when we resolve the turbulent scale height, even though we obtain a distribution of column densities sufficiently wide such that there are a small fraction of super-Eddington columns, the prospects for this mechanism to succeed in nature will depend on a more detailed examination of this possible effect in simulations with more highly supersonic turbulence and with a finite escape velocity. The idealised plane-parallel setup we use for these simulations, although it is also one of the configurations contemplated in \cite{TK_2016}, may be problematic for driving winds because the escape velocity is formally infinite. As noted by \cite{TK_2016}, a key assumption of their model is that the column density distribution remains correlated for time scales long enough that low-column patches that are locally super-Eddington have time to be accelerated by the radiation force to a significant speed before the turbulent pattern shifts and they are shadowed by opaque, sub-Eddington regions at lower altitude. However, this assumption can never be satisfied in a truly plane-parallel system, because the velocity required to escape is infinity, and thus the column density distribution would need to remain correlated for an indefinitely long time to allow material to escape to arbitrary height. In this respect the truly plane-parallel situation represents a singular limit whose behaviour may be significantly different than the case of a disc of material that is thin but has a finite scale height as a result of turbulent motions driven by supernovae, gravitational instability, or some other mechanism. We leave investigation of this case to future work.

\subsection{Dimensional limitations}
Due to the availability of computational resources, we did not perform three-dimensional simulations. However, we expect that the existence of the instability and the identified region of instability in the $( \Gamma_{\text{Edd}}, \tau)$ parameter space will not be altered by dimensionality.  These features appear to be well predicted in our two-dimensional simulations by a quasi-1D perturbative analysis (section \ref{section:perturbation_theory}) and we expect these basic features to persist in three dimensions.

However, the morphology and nonlinear development of the instability may be quite different in three dimensions, due to the existence of vortex stretching terms in the three-dimensional vorticity equation (which do not exist in the 2D vorticity equation).  We expect, analogous to the ordinary Raleigh-Taylor instability, that turbulent motions in three dimensions will lead to smaller-scale features rather than the large-scale plumes and channels we observe in two dimensions, which may affect the global behavior.  Such qualitatively different morphology in 3D compared to 2D was observed by \cite{Davis_2014} in their simulations of radiation-pressure-driven winds in the multiple-scattering limit, also finding a consistently higher volume-averaged Eddington ratio in 3D compared to 2D. We leave exploration of 2D vs. 3D in the single-scattering limit to future work.

\subsection{Thermodynamic limitations and dust-gas coupling assumptions}
We employed an isothermal equation of state in this work, assuming that the gas cooling and heating timescales are much smaller than the dynamical timescales we simulate.  However, the FUV photons providing momentum to the gas also provide thermal energy via photoelectric heating, which (assuming ionization equilibrium) could heat the gas to thousands of Kelvins in near-Eddington radiation environments, depending on the local density and the metallicity and gravitational potential of the system in question.  However, this effect has been extensively studied in simulations of disk galaxies (e.g., \citealt{Tasker_2011,Forbes_2016}) and here we seek to isolate the dynamical effects of radiation pressure. Therefore we do not make any conclusions about the thermal state of the gas (or dust) in this work, and we anticipate future work incorporating self-consistent heating and cooling source terms.

We also neglect any relative velocity between the dust and gas. The dynamics of a two-component dust and gas medium has been previously computed in spherical symmetry in the case of stellar winds \citep{Tielens_1983,Berruyer_1983,Dominik_1989,Netzer_1993}, where the two-fluid results are qualitatively similar to those obtained assuming perfect coupling, except for a very small region near the base of the wind. The dynamics of dust as aerodynamic particulates  in a supersonically turbulent interstellar medium has been considered by several authors \citep{Hopkins_2016b,Squire_2017a,Hopkins_2017,Lee_2017,Squire_2017b}. They find that small ($\sim$nm) dust grains, expected to absorb most UV/optical photons, are tightly coupled to the gas by drag forces and are even more tightly coupled in the presence of Lorentz forces, although these works neglect radiative acceleration. For the effects of radiative acceleration on dust in spherical symmetry, see \cite{Suttner_1999}.

\section{Conclusions}
\label{section:conclusion}
We have conducted a detailed investigation of the stability properties of radiation-supported two-dimensional isothermal atmospheres in the single-scattering limit both perturbatively and with simulations using our newly-developed angle-resolving radiation transport code based on the method of \cite{Reed_1973} and \cite{Klein_1989} (Appendix \ref{appendix:DG}).

We identify a small region of $(\Gamma_{\text{Edd}}, \tau)$ parameter space that is unstable and produces turbulence in the nonlinear regime (Figures \ref{fig:perturbative_stability} \& \ref{fig:stability}). Unstable solutions produce statistically steady state turbulent atmospheres with Mach numbers $\sim0.5-2$ over the range of parameters explored (Figures \ref{fig:turb_dv_all}, \ref{fig:tau_mach_scaling}, \& \ref{fig:summary_dv}). The turbulent kinetic energy density of the motions is in general substantially less than the energy density of the driving radiation field (Figures \ref{fig:erad_time} \& \ref{fig:summary_ke}). Due to both the turbulence driven by radiation pressure and the reduction in the effective gravitational acceleration, the scale height of the atmosphere is inflated to of order $\tau \Gamma_{\text{Edd}} c_T^2 / g (1 - \Gamma_{\text{Edd}})$ (Figure \ref{fig:density_profiles}, section \ref{subsection:summary}). An extrapolation of our results to extremely high optical depth ($\tau \sim 1000$) and Eddington ratio ($\Gamma_{\text{Edd}} \sim 1$) suggests that highly supersonic turbulence ($\mathcal{M} \sim 20$) could be driven by systems in this extreme of parameter space. We show that very high spatial resolution ($\lesssim 0.5$ pc for $300$ Kelvin gas in a $100 \, M_{\odot}$ pc$^{-2}$ disk) is required to resolve the instability that produces this turbulence (Appendix \ref{appendix:resolution}).

Within the idealized nature of our calculations, these results, combined with the parameter space identified by unresolved observations of star-forming galaxies (Figure \ref{fig:bimodal_aco_galaxies}), imply that radiation pressure in the single-scattering limit is not a significant contributor to supersonic turbulence in star-forming galaxies when averaged over galaxy-wide scales. However, resolved observations should indicate whether the star-forming sub-regions of such galaxies (e.g., giant molecular clouds or star clusters) are closer to the single-scattering Eddington limit and thus lie in the unstable parameter space where significant supersonic turbulent motions should be driven by the instability we identify in this work.

For galaxies substantially below the Eddington limit, we note that the column-density-dependent nature of the acceleration by radiation in a regime where the column density variations are produced by an \emph{external} turbulent driving mechanism (e.g., supernovae or gravitational instability in a multiphase medium) may yet produce mass loss and drive galactic winds, as proposed by \cite{TK_2016}. Because galaxies with low average optical depths lie preferentially closer to the Eddington limit in Figure \ref{fig:bimodal_aco_galaxies}, they may be the most susceptible to this mechanism for mass loss. We suggest that future work investigate, via controlled numerical experiments, the effectiveness of radiative acceleration in the presence of externally-driven turbulence for driving galactic winds in star-forming galaxies in the single-scattering limit.

\section*{Acknowledgements}

BDW thanks E. Ostriker and J. Stone for useful discussions at the 2016 Prospects in Theoretical Physics summer school held at the Institute for Advanced Study and C.S. Kochanek for a thorough reading of the manuscript.

BDW is supported by the National Science Foundation Graduate Research Fellowship Program under Grant No. DGE-1343012. TAT is supported by the National Science Foundation under Grant No. AST-1516967. Any opinions, findings, and conclusions or recommendations expressed in this material are those of the author(s) and do not necessarily reflect the views of the National Science Foundation. MRK acknowledges support from the Australian Research Council's Discovery Projects funding scheme (project DP160100695).

Some computations in this paper were run on the CCAPP condo of the Ruby Cluster at the Ohio Supercomputer Center \citep{Center_1987}.

We gratefully acknowledge the use of the \textsc{Matplotlib} software package \citep{Hunter_2007}. This research has made use of NASA's Astrophysics Data System.



\bibliographystyle{mnras}
\bibliography{bibliography/biblio} 



\appendix
\section{Angular quadrature}
\label{appendix:angles}
We can compute the radiation field very accurately with only \emph{one} poloidal angle (i.e., the `$f_{zz} = 1/3$' angular quadrature in the Appendix of \citealt{Davis_2014}) corresponding to the direction cosine one would obtain from two-point Gaussian quadrature:
\begin{align}
&\cos \phi = \frac{1}{\sqrt{3}} \\
&\phi = \cos^{-1} \left( \frac{1}{\sqrt{3}} \right) \approx 0.9553166183 \approx 54.735610326\deg \, .
\end{align}

As \cite{Davis_2014} note, this angular quadrature forces the $f_{zz}$ component of the Eddington tensor to be equal to $1/3$, but given this approximation, it provides the best angular resolution in 2D plane-parallel geometry for a fixed number of angles. Quadrature in the other angular coordinate $\theta$ is uniform in angle.


\section{Discontinuous Galerkin radiation transport}
\label{appendix:DG}

There are many schemes for computing radiation transport. It is a difficult problem with no single method dominating in practice over the others.  This paper discusses a state-of-the-art method for deterministic (i.e., not Monte Carlo) transport of radiation from \emph{diffuse} sources (which may either lie inside the computational volume or on the boundary).  Point sources present their own problems caused by the numerical diffusivity inherent in most deterministic transport schemes and are not treated here (although an adaptive angular quadrature scheme may make the methods discussed here viable for treating such sources).

The method discussed here is that of \emph{discontinuous Galerkin finite elements} (DGFEM), a relatively unknown methodology within computational astrophysics.  In this method, we use an approximate representation of the solution on localized basis functions and integrate by parts the continuous equation we wish to approximate. This method was originally introduced as a neutron transport method by \cite{Reed_1973}, and later extended to photon transport in the astrophysical context by \cite{Klein_1989}, with later development by \cite{Castor_1992} and \cite{Dykema_1996}. We base our derivation on that of \cite{Castor_1992} and \cite{Dykema_1996}.

This scheme is similar to the method of short characteristics, and they have the same asymptotic scaling of computational complexity in the number of spatial zones and angular ordinates.  However, there are important differences.  DGFEM does not require the exact (formal) solution along rays and the associated evaluations of exponentials, which can still be quite expensive on modern CPUs and GPUs. Second, it does not require any scheme for interpolating the upwind specific intensities, which virtually eliminates the problems with (spurious) negative values of the specific intensity, which otherwise require expensive `limiters' of the sort used in hydrodynamic solvers to maintain positivity of the solution (e.g., \citealt{Balsara_2001}). These advantageous properties come at the expense of maintaining in memory solution weights for each basis function in each spatial zone.

\subsection{Equations}
\subsubsection{Continuous transport equation}
For a given angle (omitted in the notation), we use the transport equation in its standard form, neglecting differences between the lab and comoving frames:
\begin{align}
\frac{1}{c} \frac{\partial I}{\partial t} + \hat{\Omega} \cdot \vec{\nabla} I = \eta - \chi I \, .
\end{align}

\subsubsection{Discrete transport equation}
For a given angle (omitted in the notation), within a given element, the intensity is represented as the product of the element basis functions and the nodal weights $I_l$, as such:
\begin{align}
I(x,y) = \sum_{l = 1}^{N} w_l (x,y) I_l \, .
\end{align}

Integrating by parts and carefully treating where the delta functions yield fluxes between finite elements, we obtain a discrete transport equation, for $I$ as the downwind specific intensity, $I^{*}$ as the upwind specific intensity, $\hat{\Omega}$ as the angle, and $\hat{n}$ as the outward-pointing normal of the downwind element on the boundary between the upwind and downwind elements (such that $\hat{\Omega} \cdot \hat{n} < 0$):
\begin{equation}
\begin{split}
\sum_{k = 1}^{N} \left[ \left( \frac{1}{c} \frac{\partial I_k}{\partial t} + \chi I_k - \eta_k \right) \int dV w_l w_k - I_k \hat{\Omega} \cdot \int dV w_k \vec{\nabla} w_l \right. \\
\left. + I_k \int_{\partial^{+} V} dA w_l w_k \hat{\Omega} \cdot \hat{n} + I_k^{*} \int_{\partial^{-} V} dA w_l w_k \hat{\Omega} \cdot \hat{n} \right] = 0
\end{split}
\end{equation}
for $l = 1, \ldots, N$.

Approximating the integral $\int dV w_l w_k$ as $\delta_{kl} \int dV w_l$ (`mass lumping' in the terminology of \citealt{Castor_1992}) yields:
\begin{equation}
\begin{split}
\left( \frac{1}{c} \frac{\partial I_l}{\partial t} + \chi I_l - \eta_l \right) \int dV w_l \\
+ \sum_{k=1}^{N} \left( -I_k \hat{\Omega} \cdot \int dV w_k \vec{\nabla} w_l \right. \\
\left. + I_k \int_{\partial^{+} V} dA w_l w_k \hat{\Omega} \cdot \hat{n} + I_k^{*} \int_{\partial^{-} V} dA w_l w_k \hat{\Omega} \cdot \hat{n} \right) = 0
\end{split}
\end{equation}
\begin{equation}
\left( \frac{1}{c} \frac{\partial I_l}{\partial t} + \chi I_l - \eta_l \right) V_l + \sum_{k=1}^{N} \left( I_k D_{k,l} + I_k^{*} D^{*}_{k,l} \right) = 0
\end{equation}
\begin{equation}
D_{k,l} =  -\hat{\Omega} \cdot \int dV w_k \vec{\nabla} w_l + \int_{\partial^{+} V} dA \,w_k w_l \,\hat{\Omega} \cdot \hat{n}
\end{equation}
for $l = 1, \ldots, N$.  $\int_{\partial V^{-}}$ denotes an integral over the upwind boundary of the element (for which $\hat{\Omega} \cdot \hat{n} < 0$), while $\int_{\partial V^{+}}$ denotes an integral over the downwind boundary of the element (for which $\hat{\Omega} \cdot \hat{n} > 0$).

\subsection{First-order scheme}
If we choose the simplest possible (i.e., constant) basis for our finite elements, then this equation becomes:
\begin{equation}
\left( \frac{1}{c} \frac{\partial I}{\partial t} + \chi I - \eta \right) \int dV + \left( I \int_{\partial^{+} V} dA \, \hat{\Omega} \cdot \hat{n} + I^{*} \int_{\partial^{-} V} dA\, \hat{\Omega} \cdot \hat{n} \right) = 0
\end{equation}
Assuming a regular Cartesian mesh:
\begin{equation}
\begin{split}
\frac{1}{c} \frac{\partial I}{\partial t}\, \Delta x \Delta y + \left( \chi\, \Delta x \Delta y + \Delta y \, \hat{\Omega}_x + \Delta x \, \hat{\Omega}_y \right) I \\
= \eta\, \Delta x \Delta y +  \left( I^{*}_x \Delta y \, \hat{\Omega}_x + I^{*}_y \Delta x \, \hat{\Omega}_y \right)
\end{split}
\end{equation}
Using a backward-Euler finite difference in time:

\begin{equation}
\begin{split}
\frac{1}{c \Delta t} \left( I - I^{t-1} \right) \, \Delta x \Delta y + \left( \chi\, \Delta x \Delta y + \Delta y \, \hat{\Omega}_x + \Delta x \, \hat{\Omega}_y \right) I \\
= \eta\, \Delta x \Delta y +
 I^{*}_x \Delta y \, \hat{\Omega}_x + I^{*}_y \Delta x \, \hat{\Omega}_y
\end{split}
\end{equation}
\begin{equation}
\begin{split}
\left( \frac{1}{c \Delta t} \, \Delta x \Delta y + \chi\, \Delta x \Delta y + \Delta y \, \hat{\Omega}_x + \Delta x \, \hat{\Omega}_y \right) I \\
= \eta\, \Delta x \Delta y + I^{*,x} \Delta y \, \hat{\Omega}_x + I^{*,y} \Delta x \, \hat{\Omega}_y
 + \frac{1}{c \Delta t} I^{t-1}\, \Delta x \Delta y
\end{split}
\end{equation}
\begin{equation}
I = \frac{\eta\, \Delta x \Delta y + I^{*,x} \Delta y \, \hat{\Omega}_x + I^{*,y} \Delta x \, \hat{\Omega}_y  + \frac{1}{c \Delta t} I^{t-1} \, \Delta x \Delta y }{\left( \frac{1}{c \Delta t} \, \Delta x \Delta y + \chi\, \Delta x \Delta y + \Delta y \, \hat{\Omega}_x + \Delta x \, \hat{\Omega}_y \right)}
\end{equation}
\begin{equation}
I = \frac{\eta + I^{*,x} \frac{\hat{\Omega}_x}{\Delta x} + I^{*,y}\frac{\hat{\Omega}_y}{\Delta y} + \frac{1}{c \Delta t} I^{t-1} }{\left( \frac{1}{c \Delta t} + \chi + \frac{ \hat{\Omega}_x}{\Delta x} + \frac{\hat{\Omega}_y}{\Delta y} \right)}.
\end{equation}
So for first-order DG transport, the recursive update rule for the (downwind) specific intensity can be expressed in closed form without any matrix inversions.  This method can be expected to converge linearly in space and time, and as such is somewhat numerically diffusive.

However, the first-order method is simple to code and adequate for computing transport in the semitransparent regime.  We therefore adopt it for the simulations in this paper.  The derived scheme is exactly equivalent to the finite difference scheme of \cite{Stenholm_1991}, although here we use finite \emph{elements}, rather than finite differences, for the derivation.

\subsection{Time discretization}
We discretize the time derivative with a backward Euler finite difference representation:

\begin{equation}
\frac{1}{c} \frac{\partial I}{\partial t} \approx \frac{1}{c} \frac{I(t_0 + \Delta t) - I(t_0)}{\Delta t} \approx \frac{1}{c \Delta t} \left( I_{t} - I_{t-1} \right).
\end{equation}
Then an initial-value problem of the form
\begin{equation}
\frac{1}{c} \frac{\partial I}{\partial t} = C
\end{equation}
becomes a boundary value problem of the form
\begin{equation}
I_{t}  = c \Delta t\, C + I_{t-1}.
\end{equation}

\subsection{Transport sweeps}
We invert the discretized transport equation element-by-element, starting with the most-upwind element for a given angle, inverting the local ($N$x$N$, where $N$ is the number of basis functions per element) transport matrix to obtain the $I_k$'s for that element, then proceeding with all downwind elements in sequence.

For a serial algorithm, it is simplest to choose the next element to compute as the rightmost element nearest to the present element along one axis, and exhausting the elements along that row.  Then we move to the next-downwind row and sweep laterally along that row. For a parallel algorithm, the sweep may be taken diagonally, instead of laterally \citep{Koch_1992}.

\subsection{Cell-averaged moments}
The cell-averaged energy density $J$ is given by
\begin{equation}
\begin{split}
J = \frac{1}{n \Delta V}\sum_{n}  \int dV I_n(a,b) = \frac{1}{n \Delta V}\sum_{n}  \int dV \sum_{l = 1}^{N} w_l (x,y) I_{n,l} \\
= \frac{1}{n \Delta V}\sum_{n}  \sum_{l = 1}^{N} I_{n,l} \int dV  w_l (x,y) = \frac{1}{n}\sum_{n}  \sum_{l = 1}^{N} I_{n,l} V_l \, .
\end{split}
\end{equation}

The cell-averaged flux density $F$ is given by
\begin{align}
F_{\hat{n}} &= \frac{1}{n \Delta V}\sum_{n}  \int dV I_n(a,b) \, \hat{\Omega}_n \cdot \hat{n} \nonumber \\
&= \frac{1}{n \Delta V}\sum_{n}  \int dV \sum_{l = 1}^{N} w_l (x,y) I_{n,l} \, \hat{\Omega}_n \cdot \hat{n} \nonumber \\
&= \frac{1}{n \Delta V}\sum_{n}  \sum_{l = 1}^{N} I_{n,l} \, \hat{\Omega}_n \cdot \hat{n}  \int dV  w_l (x,y) \nonumber \\
&= \frac{1}{n}\sum_{n}  \sum_{l = 1}^{N} I_{n,l} V_l \,\,\hat{\Omega}_n \cdot \hat{n} \, .
\end{align}

\subsection{Convergence properties}
The resulting transport algorithm is similar in many respects to the short characteristics version of discrete ordinates transport, but we can avail ourselves of the consistency and convergence theorems of discontinuous Galerkin finite element theory. Since we used zeroth-order (constant) basis functions, the method converges at first-order in spatial resolution.

When using first-order (piecewise linear) basis functions, this method yields the correct diffusion limit (i.e., accurately reproduces the first and second moments of the specific intensity in an asymptotic expansion at $\tau \rightarrow \infty$) for the energy and flux of the radiation field \citep{Castor_1992}.


\section{Resolution tests}
\label{appendix:resolution}
To check the robustness of our numerical conclusions about stability, we have conducted a number of tests to explore the sensitivity of our results to resolution and box size. With respect to box size, we find that there is a critical horizontal box size below which turbulent motions are greatly suppressed.  Simulations B, C, D, and E (in Table \ref{table:sims}) have horizontal box sizes that were chosen to be close to the minimum unstable wavelength for a $\tau = 10$, $\Gamma_{\text{Edd}} = 0.8$ system derived in section \ref{section:perturbation_theory}. These simulations appear as the points at $\mathcal{M} \sim 1$, $\tau = 10$, and $\Gamma_{\text{Edd}} = 0.8$ in Figure \ref{fig:turb_dv_all}. All have turbulent velocities reduced by a factor $\sim 2$ compared to the fiducial simulation A, which is identical except for the horizontal box size and variation of boundary condition (outflow or reflecting upper horizontal boundary).

\begin{figure}
  \includegraphics[width=\columnwidth]{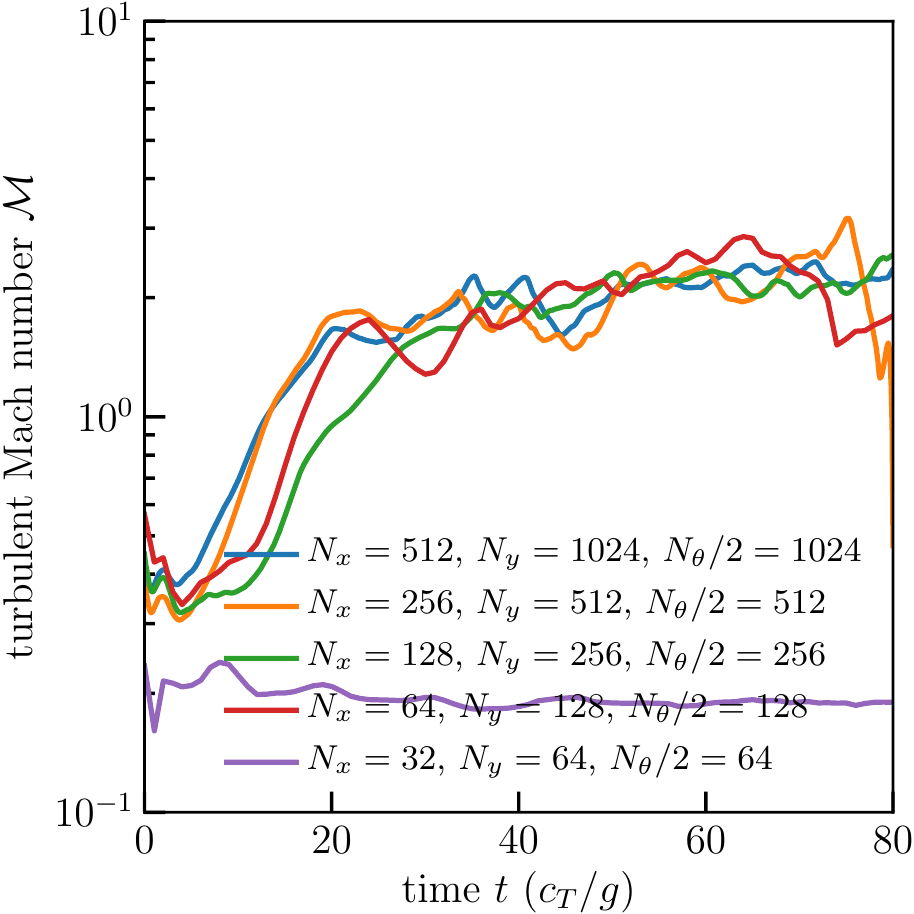}
  \caption{The turbulent velocity dispersion as a function of time for simulation A (solid blue), simulation A0 (solid orange), simulation A1 (solid green), simulation A2 (solid red), and simulation A3 (solid purple; ordered from high resolution to low resolution). The mean turbulent velocity dispersion is qualitative unaffected by changes in the spatial and angular resolution of our simulations as long as the gas pressure scale height is resolved with at least two zones.}
  \label{fig:resolution_turb_dv}
\end{figure}
To verify that our conclusions are unaffected by changes in spatial and angular resolution, we run a simulation equivalent to simulation A, but at a spatial and angular resolution twice as coarse (simulation A0), and we also run simulations that are four times as coarse (simulation A1), eight times as coarse (simulation A2), and sixteen times as coarse (simulation A3). Simulation A1 has the same resolution as the lowest-resolution simulations used in this paper. We sample lower resolutions to justify the lowest resolution we use. We show the turbulent velocity dispersion for simulation A (solid blue line), simulation A0 (solid orange line), simulation A1 (solid green line), simulation A2 (solid red line), and simulation A3 (solid purple line) in Figure \ref{fig:resolution_turb_dv}. Except for the lowest-resolution simulation, the time-dependent behavior is qualitatively similar across all simulations and there is no significant difference in time-averaged turbulent Mach number after an initial transient. (The relatively high velocity dispersions of simulation A2 are due to a rapid mass loss from the computational domain that proceeds much slower in the higher-resolution simulations.)

The lowest-resolution simulation has a very sub-sonic velocity dispersion that is slowly decaying from the initial ($t=0$) velocity perturbation. This indicates that the instability is not resolved at this low resolution. Simulation A3 has a spatial resolution
\begin{align}
\Delta x = 0.78 \, c_T^2/g = 0.72 \text{ pc} \left( \frac{T}{300 \, K} \right) \left( \frac{\Sigma}{100 \, M_{\odot} \text{ pc}^{-2}} \right)^{-1} \, .
\end{align}
The minimum resolution to resolve the instability (at our fiducial optical depth $\tau = 10$ and Eddington ratio $\Gamma_{\text{Edd}} = 0.8$) therefore lies between the resolution of simulation A2 ($\Delta x = 0.39 \, c_T^2/g$) and simulation A3 ($\Delta x = 0.78 \, c_T^2/g$), suggesting that resolving the gas pressure scale height with at least two zones is necessary to resolve the instability. We have not explored in detail whether this resolution requirement is a function of optical depth.

In a simulation with optically-thin radiative cooling, this convergence test suggests that better than 1 pc resolution would be necessary to resolve the instability. We note that effective resolution can be highly dependent on the details of the numerical method. We use high-order methods for hydrodynamics, but low-order methods for transport and for coupling the source terms; the use of accurate high-order methods for all steps of the calculation may yield less stringent resolution requirements, but constructing and testing such a code is nontrivial.


\bsp	
\label{lastpage}
\end{document}